\documentclass[aps,pre,preprint,bibnotes]{revtex4}

\usepackage{amsmath,amssymb,amsfonts,graphicx}
\usepackage{array}

\begin{document}

\title{Solving the Navier-Lam\'{e} Equation in Cylindrical Coordinates Using the Buchwald Representation: \\ Some Parametric Solutions with Applications}

\author{Jamal Sakhr}
\author{Blaine A. Chronik}
\affiliation{Department of Physics and Astronomy, The University of Western Ontario, London, Ontario N6A 3K7 Canada}

\date{\today}

\begin{abstract}
Using a separable Buchwald representation in cylindrical coordinates, we show how under certain conditions the coupled equations of motion governing the Buchwald potentials can be decoupled and then solved using well-known techniques from the theory of PDEs. Under these conditions, we then construct three parametrized families of particular solutions to the Navier-Lam\'{e} equation in cylindrical coordinates. In this paper, we specifically construct solutions having $2\pi$-periodic angular parts. These particular solutions can be directly applied to a fundamental set of linear elastic boundary value problems in cylindrical coordinates and are especially suited to problems involving one or more physical parameters. As an illustrative example, we consider the problem of determining the response of a solid elastic cylinder subjected to a time-harmonic surface pressure that varies sinusoidally along its axis, and we demonstrate how the obtained parametric solutions can be used to efficiently construct an exact solution to this problem. We also briefly consider applications to some related forced-relaxation type problems. 
\end{abstract}

\maketitle

\section{Introduction}

The Navier-Lam\'{e} (NL) equation is the fundamental equation of motion in classical linear elastodynamics \cite{Elastobuch}. This vector equation, which governs the classical motions of a homogeneous, isotropic, and linearly elastic solid, cannot generally be solved using elementary methods (such as separation of variables). Many special methods have however been developed that have proved useful for obtaining particular solutions. One of the most well-known and effective methods for generating solutions is the method of potentials, whereby the displacement vector is represented as a combination of one or more scalar and/or vector functions (called potentials). The scalar components of the NL equation constitute a set of coupled PDEs that cannot (except in rare cases) be solved exactly regardless of the coordinate system used. Representations in terms of displacement potentials may, once substituted into the NL equation, yield a system of PDEs that are (in the best case scenario) uncoupled, less coupled, or at least simpler (e.g., decreased order in the derivatives) than the original component equations. The most theoretically well-established representations include the Helmholtz-Lam\'{e}, Kovalevshi-Iacovache-Somigliana, and (dynamic) Papkovich-Neuber representations \cite{Gurtin,Miky,Ruskie}. Despite having theoretically robust properties, these classical representations can be highly cumbersome when used as analytical tools for solving problems. Other representations have been more recently proposed in studies of anisotropic media \cite{TI1,TI2}. While these representations can be specialized to isotropic media, their analytical efficacy when applied to isotropic problems remains largely unexplored.

One representation that has proven to be analytically effective in anisotropic problems with cylindrical symmetry is the so-called Buchwald representation. This representation was first proposed in 1961 in a paper on the subject of Rayleigh waves in a transversely isotropic medium \cite{Buch61}. (An interesting historical sidebar is that cylindrical coordinates were not actually used in this paper.) A number of studies published in the following decades made use of Buchwald's representation expressed in cylindrical coordinates \cite{Kundu92,Raja93}, but it was not until the late 1990s that the analytical advantages of doing so were first demonstrated in a series of papers by Ahmad and Rahman \cite{AR98,AR00,AR01}. The systems of interest in the aforementioned papers (and indeed in almost all papers employing the Buchwald representation) are transversely isotropic. The Buchwald representation is also applicable to problems involving isotropic media \cite{thesis2,ArXpap}.  Application of the Buchwald representation to such problems is however rare \cite{thesis1,Indguys1,Indguys2}. 

The Buchwald representation involves three scalar potential functions (see Eq.~(\ref{BuchyALL}) of Sec.~\ref{EOMsALL}). As shown in Ref.~\cite{thesis1} (and by different means in Sec.~\ref{EOMsALL} of this paper), the Buchwald representation reduces the original scalar components of the NL equation (which are a set of three coupled PDEs) to a set of two coupled PDEs involving two of the potentials and one separate decoupled PDE involving the remaining potential. By assuming separable product solutions and imposing certain conditions on the axial and temporal parts of the three potentials, we show that the coupled subsystem reduces to a homogeneous linear system, which can be solved by elimination. The elimination procedure generates two independent fourth-order linear PDEs with constant coefficients that can subsequently be solved using fundamental solutions to the two-dimensional Helmholtz equation in polar coordinates. We then construct particular solutions possessing $2\pi$-periodic angular parts, and care is taken to consider all allowed parameter values including those yielding degenerate cases. A simultaneous solution to the originally decoupled PDE is also obtained using separation of variables. All three Buchwald potentials can thus be completely determined (under the stipulated conditions) and what follows is the construction of three distinct families of parametric solutions to the NL equation. These solutions again possess $2\pi$-periodic angular parts. The construction of particular solutions not having $2\pi$-periodic angular parts poses additional complications and will be considered elsewhere. 

Independent of the fact that they were derived using Buchwald displacement potentials, these particular solutions to the NL equation are interesting and useful in their own right. They can be directly applied to a fundamental set of linear elastic boundary value problems in cylindrical coordinates and are especially suited to problems involving one or more physical parameters. In the second half of the paper, we focus on how these parametric solutions can be used to solve certain types of boundary-value problems. As an illustrative example, we consider the problem of determining the response of a solid elastic cylinder subjected to a time-harmonic surface pressure that varies sinusoidally along its axis. To demonstrate the range of applicability of the obtained parametric solutions, we also briefly consider applications to some related forced-relaxation type problems.  

\section{Equations of Motion}\label{EOMsALL} 

The NL equation can be written in vector form as \cite{Elastobuch,Ruskie}
\begin{subequations}\label{NLE} 
\begin{equation}\label{NLE1}
\mu\nabla^2\mathbf{u}+(\lambda+\mu)\nabla(\nabla\cdot\mathbf{u})+\mathbf{b}=\rho{\partial^2\mathbf{u}\over\partial t^2},
\end{equation}
or (using the well-known identity for the vector Laplacian \footnote{$\displaystyle \nabla^2\mathbf{u}=\nabla(\nabla\cdot\mathbf{u})-\nabla\times(\nabla\times\mathbf{u})$}) as 
\begin{equation}\label{NLE2}
(\lambda+2\mu)\nabla(\nabla\cdot\mathbf{u})-\mu\nabla\times(\nabla\times\mathbf{u})+\mathbf{b}=\rho{\partial^2\mathbf{u}\over\partial t^2},
\end{equation}
\end{subequations} 
where $\mathbf{u}$ is the displacement vector, $\lambda$ and $\mu>0$ are the Lam\'{e} constants from the classical linear theory of elasticity, $\mathbf{b}$ is the local body force, and $\rho>0$ is the (constant) density of the material. (Note that the first Lam\'{e} constant $\lambda$ need not be positive.) In this paper,  we consider the situation in which there are only surface forces acting on the solid and no body forces (i.e., $\mathbf{b}=0$). In the circular cylindrical coordinate system, the displacement $\mathbf{u}\equiv\mathbf{u}(r,\theta,z,t)$ depends on the spatial coordinates $(r,\theta,z)$ and on the time $t$. 

The Buchwald representation of the displacement field involves three scalar potentials $\Phi(r,\theta,z,t)$, $\Psi(r,\theta,z,t)$, $\chi(r,\theta,z,t)$, and can be written as \cite{AR98}
\begin{subequations}\label{BuchyALL}
\begin{equation}\label{Buchy}
\mathbf{u}=\nabla\Phi+\nabla\times(\chi\mathbf{\hat{z}})+\left({\partial\Psi\over\partial z}-{\partial\Phi\over\partial z}\right)\mathbf{\hat{z}},
\end{equation}
where $\mathbf{\hat{z}}$ is the unit basis vector in the $z$-direction. Note that many variations of the above form can be found in the literature. For example, the form  
\begin{equation}\label{Buchy2}
\mathbf{u}=-\nabla\Phi+\nabla\times(\chi\mathbf{\hat{z}})+\left({\partial\Psi\over\partial z}+{\partial\Phi\over\partial z}\right)\mathbf{\hat{z}},
\end{equation}
is used in Ref.~\cite{Sharma01}, and the form  
\begin{equation}\label{Buchy3}
\mathbf{u}=-\nabla\Phi+\nabla\times(\chi\mathbf{\hat{z}})-\left({\partial\Psi\over\partial z}-{\partial\Phi\over\partial z}\right)\mathbf{\hat{z}}.
\end{equation}
\end{subequations}
is used in Refs.~\cite{Indguys1,Indguys2,Sharma02}. Other more significant variations of the above forms that one might better refer to as ``Buchwald-type'' representations are ubiquitous in the literature (see, for example, Ref.~\cite{Chen99}). The form (\ref{Buchy}) advocated by Ref.~\cite{AR98} is however most common and will be employed here. Although not relevant for our purposes, we should mention that the completeness of the Buchwald representation has not (to our knowledge) been rigorously proven. Some authors do however refer to it as being a complete representation \cite{BuchComp}. 

When Eq.~(\ref{Buchy}) is substituted into Eq.~(\ref{NLE}), the resulting equation can (after some manipulation) be put into the form
\begin{subequations}\label{NLBE}
\begin{equation}\label{NLBE1}
\nabla\left[(\lambda+2\mu)\nabla^2\Phi+(\lambda+\mu)(\nabla\cdot{\boldsymbol{\Upsilon}})-\rho{\partial^2\Phi\over\partial t^2}\right] + \nabla\times \left[\mu\nabla^2{\boldsymbol{\chi}}-\rho{\partial^2\boldsymbol{\chi}\over\partial t^2}\right] + \left[\mu\nabla^2{\boldsymbol{\Upsilon}}-\rho{\partial^2\mathbf{\Upsilon}\over\partial t^2}\right]=0, 
\end{equation}
where 
\begin{equation}\label{NLBE2}
\boldsymbol{\Upsilon}=\left({\partial\Psi\over\partial z}-{\partial\Phi\over\partial z}\right)\mathbf{\hat{z}},
\end{equation}
and
\begin{equation}\label{NLBE3}
\boldsymbol{\chi}=\chi\mathbf{\hat{z}}.
\end{equation}
\end{subequations}
The Buchwald representation of the displacement vector field satisfies Eq.~(\ref{NLE}) identically (when $\mathbf{b}=0$ \footnote{For the general case $\mathbf{b}\neq0$, one must determine a suitable representation of $\mathbf{b}$ in terms of one or more scalar and/or vector potentials called body-force potentials. These body-force potentials will generally differ from the displacement potentials that represent $\mathbf{u}$. The set of displacement and body-force potentials representing $\mathbf{u}$ and $\mathbf{b}$, respectively, must each satisfy certain conditions and be compatible so as to satisfy the NL equation. While such representations of $\mathbf{b}$ exist and are well-known for the well-established representations of $\mathbf{u}$ \cite{Gurtin,Miky,Pak07}, representations of $\mathbf{b}$ in terms of potentials compatible with the Buchwald representation of $\mathbf{u}$ have not yet been put forward in the literature.}) provided $\Phi,\Psi$, and $\chi$ are solutions of the coupled system
\begin{subequations}\label{systemeB} 
\begin{equation}
(\lambda+2\mu)\nabla^2\Phi+(\lambda+\mu)(\nabla\cdot{\boldsymbol{\Upsilon}})-\rho{\partial^2\Phi\over\partial t^2}=0,
\end{equation}
\begin{equation}
\mu\nabla^2{\boldsymbol{\Upsilon}}-\rho{\partial^2\mathbf{\Upsilon}\over\partial t^2}=0,
\end{equation}
\begin{equation}
\mu\nabla^2{\boldsymbol{\chi}}-\rho{\partial^2\boldsymbol{\chi}\over\partial t^2}=0.
\end{equation}
\end{subequations}
Note that while simultaneous solutions to system (\ref{systemeB}) identically satisfy Eq.~(\ref{NLE}), system (\ref{systemeB}) only constitutes a sufficient condition for Eq.~(\ref{NLBE1}) to hold; it does not constitute a necessary condition.

By virtue of Eqs.~(\ref{NLBE2}) and (\ref{NLBE3}), system (\ref{systemeB}) can be reduced to the scalar form:
\begin{subequations}\label{systemeS} 
\begin{equation}\label{systemeS1}
(\lambda+2\mu)\nabla^2\Phi+(\lambda+\mu){\partial^2\Psi\over\partial z^2}-(\lambda+\mu){\partial^2\Phi\over\partial z^2}-\rho{\partial^2\Phi\over\partial t^2}=0,
\end{equation}
\begin{equation}\label{systemeS2}
{\partial\over\partial z}\left(\mu\nabla^2\Psi-\mu\nabla^2\Phi-\rho{\partial^2\Psi\over\partial t^2}+\rho{\partial^2\Phi\over\partial t^2}\right)=0,
\end{equation}
\begin{equation}\label{systemeS3}
\mu\nabla^2\chi-\rho{\partial^2\chi\over\partial t^2}=0.
\end{equation}
\end{subequations}
Equation (\ref{systemeS2}) is satisfied identically (a sufficient but not necessary condition) when
\begin{subequations}\label{systemeS2P2} 
\begin{equation}\label{systemeS2P21}
\mu\nabla^2\Psi-\mu\nabla^2\Phi-\rho{\partial^2\Psi\over\partial t^2}+\rho{\partial^2\Phi\over\partial t^2}=0.
\end{equation}
Using Eq.~(\ref{systemeS1}) to substitute for the last term in Eq.~(\ref{systemeS2P21}) yields 
\begin{equation}\label{systemeS2P22}
(\lambda+\mu)\left[\nabla^2\Phi-{\partial^2\Phi\over\partial z^2}\right] + \mu\nabla^2\Psi + (\lambda+\mu){\partial^2\Psi\over\partial z^2} - \rho{\partial^2\Psi\over\partial t^2}=0. 
\end{equation}
\end{subequations}
Thus, for future reference, the PDE system of interest is the following:
\begin{subequations}\label{systemeSFIN} 
\begin{equation}\label{systemeSFIN1}
(\lambda+2\mu)\nabla^2\Phi+(\lambda+\mu){\partial^2\Psi\over\partial z^2}-(\lambda+\mu){\partial^2\Phi\over\partial z^2}-\rho{\partial^2\Phi\over\partial t^2}=0,
\end{equation}
\begin{equation}\label{systemeSFIN2}
(\lambda+\mu)\left[\nabla^2\Phi-{\partial^2\Phi\over\partial z^2}\right] + \mu\nabla^2\Psi + (\lambda+\mu){\partial^2\Psi\over\partial z^2} - \rho{\partial^2\Psi\over\partial t^2}=0. 
\end{equation}
\begin{equation}\label{systemeSFIN3}
\mu\nabla^2\chi-\rho{\partial^2\chi\over\partial t^2}=0.
\end{equation}
\end{subequations} 
Rahman and Ahmad \cite{AR98} employed decomposition (\ref{Buchy}) to obtain the equations governing the Buchwald potentials $\Phi, \Psi$, and $\chi$ in the case of a transversely isotropic solid. Equations (\ref{systemeSFIN1}), (\ref{systemeSFIN2}), and (\ref{systemeSFIN3}) above are consistent with Eqs.~(6a), (6b), and (6c) of Ref.~\cite{AR98}, respectively, when the latter are specialized to the isotropic case. 

Our initial goal is to obtain particular separable solutions to subsystem (\ref{systemeSFIN1},\ref{systemeSFIN2}) under certain conditions. To our knowledge, subsystem (\ref{systemeSFIN1},\ref{systemeSFIN2}) has never been solved directly. The standard approach to dealing with systems like (\ref{systemeSFIN}) would be to substitute cylindrical-wave solutions with undetermined coefficients into system (\ref{systemeSFIN}) and then solve for the unknown coefficients \cite{Kundu92,AR98,AR00,AR01,thesis1,Chen99}. (The aforementioned cylindrical-wave solutions are of course the common fundamental solutions to the wave equation in cylindrical coordinates.) In the following section, we pursue a different and more general approach. 

\section{Particular Solutions}\label{PartSolns} 

As a first step, we re-write subsystem (\ref{systemeSFIN1},\ref{systemeSFIN2}) in a slightly different form that makes it more amenable to obtaining separable solutions of the form:
\begin{subequations}\label{solform} 
\begin{equation}\label{solform1}
\Phi(r,\theta,z,t)=\phi_\perp(r,\theta)\phi_z(z)\phi_t(t),
\end{equation}
\begin{equation}\label{solform2}
\Psi(r,\theta,z,t)=\psi_\perp(r,\theta)\psi_z(z)\psi_t(t),
\end{equation}
\end{subequations}
which together with certain conditions on the functions $\displaystyle \{\phi_z(z),\phi_t(t),\psi_z(z),\psi_t(t)\}$ (to be specified below) define the special class of separable solutions to subsystem (\ref{systemeSFIN1},\ref{systemeSFIN2}) that we seek to find in this paper. Using the notation
\begin{equation}\label{2DLap}
\nabla_\perp^2\equiv\nabla^2-{\partial^2\over\partial z^2}={\partial^2\over\partial r^2} + {1\over r}{\partial\over\partial r} + {1\over r^2}{\partial^2\over\partial \theta^2},
\end{equation}
subsystem (\ref{systemeSFIN1},\ref{systemeSFIN2}) can be cast in the form:
\begin{subequations}\label{systemeF} 
\begin{equation}\label{systemeF1}
(\lambda+2\mu)\nabla_\perp^2\Phi+\mu{\partial^2\Phi\over\partial z^2} + (\lambda+\mu){\partial^2\Psi\over\partial z^2} - \rho{\partial^2\Phi\over\partial t^2}=0,
\end{equation}
\begin{equation}\label{systemeF2}
(\lambda+\mu)\nabla_\perp^2\Phi+\mu\nabla_\perp^2\Psi+(\lambda+2\mu){\partial^2\Psi\over\partial z^2}-\rho{\partial^2\Psi\over\partial t^2}=0.
\end{equation}
\end{subequations}
In the present context, $\nabla_\perp^2$ is the two-dimensional Laplacian operator in polar coordinates. In the context of transversely isotropic media, $\nabla_\perp^2$ is referred to as the transverse scalar Laplacian operator or the Laplacian for the transverse coordinates $r$ and $\theta$.

Substituting (\ref{solform}) into system (\ref{systemeF}) leads to the system
\begin{subequations}\label{systemesub} 
\begin{equation}\label{systemesub1}
\left[(\lambda+2\mu)\phi_z\phi_t\nabla_\perp^2+\mu\phi_z''\phi_t - \rho\phi_z\phi_t''\right]\phi_\perp + \left[(\lambda+\mu)\psi_z''\psi_t\right]\psi_\perp=0,
\end{equation}
\begin{equation}\label{systemesub2}
\left[(\lambda+\mu)\phi_z\phi_t\nabla_\perp^2\right]\phi_\perp+\left[\mu\psi_z\psi_t\nabla_\perp^2+(\lambda+2\mu)\psi_z''\psi_t-\rho\psi_z\psi_t''\right]\psi_\perp=0.
\end{equation}
\end{subequations}
A special class of solutions can be obtained directly from the method of elimination by assuming common functional dependences in the axial and temporal parts of both potentials, that is, 
\begin{subequations}\label{assump} 
\begin{equation}\label{assump3}
\phi_z(z)=\psi_z(z),
\end{equation}
\begin{equation}\label{assump6}
\phi_t(t)=\psi_t(t),
\end{equation}
and by further assuming that the derivatives of the above functions are such that
\begin{equation}\label{assump1}
\phi_z''(z)=\kappa\phi_z(z),
\end{equation}
\begin{equation}\label{assump2}
\psi_z''(z)=\kappa\psi_z(z),
\end{equation}
\begin{equation}\label{assump4}
\phi_t''(t)=\tau\phi_t(t),
\end{equation}
\begin{equation}\label{assump5}
\psi_t''(t)=\tau\psi_t(t),
\end{equation}
\end{subequations}
where $\displaystyle \kappa\in\mathbb{R}$ and $\displaystyle \tau\in\mathbb{R}\backslash\{0\}$ are free parameters. With these assumptions system (\ref{systemesub}) reduces to
\begin{subequations}\label{systemefin} 
\begin{equation}\label{systemefin1}
\left[(\lambda+2\mu)\nabla_\perp^2+\mu\kappa - \rho\tau\right]\phi_\perp + \left[(\lambda+\mu)\kappa\right]\psi_\perp=0,
\end{equation}
\begin{equation}\label{systemefin2}
\left[(\lambda+\mu)\nabla_\perp^2\right]\phi_\perp+\left[\mu\nabla_\perp^2+(\lambda+2\mu)\kappa-\rho\tau\right]\psi_\perp=0, 
\end{equation}
\end{subequations}
which has the form of a homogeneous linear system:
\begin{subequations}\label{systemefinL} 
\begin{equation}\label{systemefinL1}
L_1\phi_\perp + L_2\psi_\perp=0,
\end{equation}
\begin{equation}\label{systemefinL2}
L_3\phi_\perp+L_4\psi_\perp=0, 
\end{equation}
where the linear operators $\{L_1,L_2,L_3,L_4\}$ 
\begin{eqnarray}
L_1&=&\left[(\lambda+2\mu)\nabla_\perp^2+\mu\kappa - \rho\tau\right], \\
L_2&=&\left[(\lambda+\mu)\kappa\right], \\
L_3&=&\left[(\lambda+\mu)\nabla_\perp^2\right], \\
L_4&=&\left[\mu\nabla_\perp^2+(\lambda+2\mu)\kappa-\rho\tau\right],
\end{eqnarray}
\end{subequations}
are each first-order polynomial differential operators in $\nabla_\perp^2$ with constant coefficients. Two distinct cases can be distinguished at this point depending on whether $\kappa\neq0$ or $\kappa=0$. The former is the general case while the latter is a special case for which the linear system (\ref{systemefinL}) decouples. We shall consider these two cases separately. 

\subsection{General Case: $\displaystyle\kappa\neq0$}\label{solnkapneq0}

As stated earlier, under conditions (\ref{assump}) and assuming $\kappa\neq0$, we can solve for $\phi_\perp(r,\theta)$ and $\psi_\perp(r,\theta)$ using the method of elimination. It can be shown that the solutions $\phi_\perp(r,\theta)$ and $\psi_\perp(r,\theta)$ independently satisfy the decoupled equations: 
\begin{subequations}\label{detty} 
\begin{eqnarray}\label{detty1}
(L_1L_4-L_2L_3)\phi_\perp=0,
\end{eqnarray}
\begin{eqnarray}\label{detty2}
(L_1L_4-L_2L_3)\psi_\perp=0, 
\end{eqnarray}
\end{subequations}
where $(L_1L_4-L_2L_3)$ is the operational determinant of system (\ref{systemefinL}). By inspection, this determinant is non-zero, and thus a unique solution to system (\ref{systemefinL}) exists. We are free to solve Eq.~(\ref{detty1}) to obtain a solution for $\phi_\perp$ and then use either of Eqs.~(\ref{systemefinL1}) or (\ref{systemefinL2}) to solve for $\psi_\perp$, or alternatively, we can first solve Eq.~(\ref{detty2}) for $\psi_\perp$ and then use either of Eqs.~(\ref{systemefinL1}) or (\ref{systemefinL2}) to solve for $\phi_\perp$. (Note that these procedures do not generate spurious arbitrary constants.) For convenience, we choose the first option.  

After some straightforward algebra, it can be shown that Eq.~(\ref{detty1}) assumes the form
\begin{subequations}\label{phipde} 
\begin{equation}\label{phipde1}
\left[a_2\nabla_\perp^4+a_1\nabla_\perp^2+a_0\right]\phi_\perp=0,
\end{equation}
where
\begin{equation}\label{phipde2}
a_2=\mu(\lambda+2\mu),
\end{equation}
\begin{equation}\label{phipde3}
a_1=2\mu(\lambda+2\mu)\kappa-(\lambda+3\mu)\rho\tau,
\end{equation}
\begin{equation}\label{phipde4}
a_0=(\mu\kappa-\rho\tau)\left[(\lambda+2\mu)\kappa-\rho\tau\right].
\end{equation}
\end{subequations}
Since the operator in Eq.~(\ref{phipde1}) is a polynomial differential operator in $\nabla_\perp^2$ with constant coefficients, we may look for solutions (denoted by $\varphi$) of the same fundamental form as those of the two-dimensional Helmholtz equation in polar coordinates ($\nabla_\perp^2\varphi=\Lambda\varphi$). This leads to the characteristic equation 
\begin{equation}\label{chareq}
a_2\Lambda^2+a_1\Lambda+a_0=0,
\end{equation}
whose roots, obtained from the quadratic formula
\begin{subequations}\label{rootsc} 
\begin{equation}\label{rootsc1}
\Lambda_{\pm}={-a_1\pm\sqrt{a_1^2-4a_2a_0}\over2a_2},
\end{equation}
are
\begin{equation}\label{rootsSC}
\Lambda_-=-\left(\kappa-{\rho\tau\over(\lambda+2\mu)}\right), \quad \Lambda_+=-\left(\kappa-{\rho\tau\over\mu}\right).
\end{equation}
\end{subequations}

\vspace*{0.25cm}

If $\varphi_-(r,\theta)$ and $\varphi_+(r,\theta)$ denote the two independent general solutions to Eq.~(\ref{phipde}) corresponding to the roots $\Lambda_-$ and $\Lambda_+$, respectively, then $\phi_\perp(r,\theta)=\varphi_-(r,\theta)+\varphi_+(r,\theta)$ \footnote{$\phi_\perp(r,\theta)$ can in fact be any linear combination of $\varphi_-(r,\theta)$ and $\varphi_+(r,\theta)$, but since arbitrary constants are already built into the solutions $\varphi_-(r,\theta)$ and $\varphi_+(r,\theta)$, taking arbitrary linear combinations is redundant. Note that our method of solving the fourth-order PDE given by Eq.~(\ref{phipde1}) is based on a lesser-known result from the theory of PDEs (see, for example, Result 5 on page 1044 of Ref.~\cite{Russians2016}). One may question whether the solutions generated by this method are complete. This mathematically subtle question has, to our knowledge, not been rigorously addressed in the literature. Thus, completeness is not guaranteed. However, this is not an issue here since we only seek to find particular solutions to the PDE system (\ref{systemeSFIN}).}. 
We shall, for notational convenience, let $\phi^{(1)}_\perp(r,\theta)\equiv\varphi_-(r,\theta)$ and $\phi^{(2)}_\perp(r,\theta)\equiv\varphi_+(r,\theta)$. In this more intuitive notation, 
\begin{equation}\label{phiperpGEN}
\phi_\perp(r,\theta)=\sum_{s=1}^2\phi^{(s)}_\perp(r,\theta)=\sum_{s=1}^2\phi^{(s)}_r(r)\phi^{(s)}_\theta(\theta).
\end{equation}
The angular parts of $\varphi_-(r,\theta)$ and $\varphi_+(r,\theta)$ can, in theory, be periodic or non-periodic functions of $\theta$. The latter are pertinent in certain problems, such as those involving open cylindrical shells or panels, where periodicity in the angular coordinate is not necessary or inappropriate \cite{panelEG1,panelEG2}. \emph{In this paper, we shall consider only the class of general solutions to system (\ref{systemefinL}) that possess $2\pi$-periodic angular parts.}

\noindent When the parameters are such that both $\displaystyle (\lambda+2\mu)\kappa \neq {\rho\tau}$ and $\displaystyle \mu\kappa \neq {\rho\tau}$, particular solutions to Eq.~(\ref{phipde}) can be written as
\begin{subequations}\label{solphiP}
\begin{eqnarray}
\phi_\perp(r,\theta)=\sum_{s=1}^2\left[A_s\left \{ \begin{array}{c}
             J_n(\alpha_sr) \\
             I_n(\alpha_sr)
           \end{array} \right\} + B_s \left \{ \begin{array}{c}
             Y_n(\alpha_sr) \\
             K_n(\alpha_sr)
           \end{array} \right\}\right] \Big[C_s\cos(n\theta)+D_s\sin(n\theta)\Big],  
\end{eqnarray}
where
\begin{equation}\label{alphas}
\alpha_1=\sqrt{~\left|\kappa-{\rho\tau\over(\lambda+2\mu)}\right|~}, \quad \alpha_2=\sqrt{~\left|\kappa-{\rho\tau\over\mu}\right|~},
\end{equation}
\end{subequations}

\vspace*{0.25cm}

\noindent $\displaystyle n$ is a non-negative integer, and $\{A_1,A_2,B_1,B_2,C_1,C_2,D_1,D_2\}$ are arbitrary constants. The correct linear combination of Bessel functions in the radial part of each term in Eq.~(\ref{solphiP}) is determined by the relative values of the parameters, as given in Table \ref{TabLinCombos}. 

\vspace*{0.25cm}

\setlength{\extrarowheight}{10pt}
\begin{table}[h]
\centering
\begin{tabular}{| c | c | c |}
    \hline
    ~~~Linear Combination~~~ & ~~~~~ $s=1$ term ~~~~~ & ~~~~~$s=2$ term ~~~~~\\ [10pt] \hline
    $\displaystyle\{J_n(\alpha_sr), Y_n(\alpha_sr)\}$ & $\displaystyle \kappa > {\rho\tau\over(\lambda+2\mu)}$ & $\displaystyle \kappa > {\rho\tau\over\mu}$ \\ [10pt] \hline
    $\displaystyle\{I_n(\alpha_sr), K_n(\alpha_sr)\}$ & $\displaystyle \kappa < {\rho\tau\over(\lambda+2\mu)}$ & $\displaystyle \kappa < {\rho\tau\over\mu}$ \\ [10pt] \hline
\end{tabular}
\caption{Conditions on the radial part of each term in Eq.~(\ref{solphiP}).}
\label{TabLinCombos}
\end{table}
\setlength{\extrarowheight}{1pt}

\noindent When either $\displaystyle (\lambda+2\mu)\kappa = {\rho\tau}$ or $\displaystyle \mu\kappa = {\rho\tau}$, the constant $a_0$ defined by Eq.~(\ref{phipde4}) vanishes. The characteristic equation [Eq.~(\ref{chareq})] in either case will thus have one zero root and one non-zero root. (This fact can also be deduced from inspection of the roots given in (\ref{rootsSC}).) In obtaining (\ref{solphiP}), we initially sought out to find the particular solutions of Eq.~(\ref{phipde}) that are solutions to the 2D Helmholtz equation. The corresponding subset of particular solutions associated with the zero roots are in fact particular solutions to the 2D Laplace equation in polar coordinates. In the special case of zero roots, Eq.~(\ref{solphiP}) thus requires the following modifications.  When $\displaystyle (\lambda+2\mu)\kappa = {\rho\tau}$, the particular solution $\varphi_-(r,\theta)$ used in Eq.~(\ref{solphiP}) is no longer valid and must be replaced by a particular solution to the 2D Laplace equation. In effect, the radial part of the $s=1$ term in Eq.~(\ref{solphiP}) is incorrect and must be replaced by 
\begin{eqnarray}\label{LP1}
\phi^{(1)}_r(r)=\left \{ \begin{array}{lr}
           A_1+B_1\ln r & ~\text{if}~n=0 \\ 
           A_1r^n+B_1r^{-n} & ~\text{if}~n\neq0 \\
           \end{array} \right., \quad \text{when}~\alpha_1=0.
\end{eqnarray}
When $\displaystyle \mu\kappa = {\rho\tau}$, the particular solution $\varphi_+(r,\theta)$ used in Eq.~(\ref{solphiP}) is no longer valid and must be replaced. In this case, the radial part of the $s=2$ term in Eq.~(\ref{solphiP}) is incorrect and must be replaced by               
\begin{eqnarray}\label{LP2}
\phi^{(2)}_r(r)=\left \{ \begin{array}{lr}
           A_2+B_2\ln r & ~\text{if}~n=0 \\ 
           A_2r^n+B_2r^{-n} & ~\text{if}~n\neq0 \\
           \end{array} \right., \quad \text{when}~\alpha_2=0.
\end{eqnarray}

To obtain the solution for $\psi_\perp(r,\theta)$, it is simplest to re-arrange Eq.~(\ref{systemefin1}) and (using the above solution for $\phi_\perp(r,\theta)$) directly solve for $\psi_\perp(r,\theta)$: 
\begin{equation}\label{psi_sol_sp}
\psi_\perp(r,\theta)=-{\left[(\lambda+2\mu)\nabla_\perp^2+(\mu\kappa - \rho\tau)\right]\over(\lambda+\mu)\kappa}\phi_\perp(r,\theta).
\end{equation}
Using the facts that
\begin{equation}
\phi_\perp(r,\theta)=\varphi_-(r,\theta)+\varphi_+(r,\theta), \quad \nabla^2_\perp\phi_\perp(r,\theta)=\Lambda_-\varphi_-(r,\theta)+\Lambda_+\varphi_+(r,\theta),
\end{equation}
Eq.~(\ref{psi_sol_sp}) reduces as follows: 
\begin{eqnarray}\label{psi_sol_process}
\psi_\perp&=&-{\left[(\lambda+2\mu)\left(\Lambda_-\varphi_-+\Lambda_+\varphi_+\right)+(\mu\kappa - \rho\tau)\left(\varphi_-+\varphi_+\right)\right]\over(\lambda+\mu)\kappa} \nonumber \\
&=&-{\left[\left\{(\lambda+2\mu)\Lambda_-+(\mu\kappa - \rho\tau)\right\}\varphi_-+\left\{(\lambda+2\mu)\Lambda_++(\mu\kappa - \rho\tau)\right\}\varphi_+\right]\over(\lambda+\mu)\kappa} \nonumber \\
&=&-{\left[-(\lambda+\mu)\kappa \varphi_--(\lambda+\mu)\left(\kappa - {\rho\tau\over\mu}\right)\varphi_+\right]\over(\lambda+\mu)\kappa} \nonumber \\
&=&\varphi_-+{1\over\kappa}\left(\kappa - {\rho\tau\over\mu}\right)\varphi_+,
\end{eqnarray}
where the values of $\Lambda_\pm$ in Eq.~(\ref{rootsSC}) have been used in obtaining the third equality from the second. We have therefore shown that 
\begin{subequations}\label{solpsiP}
\begin{eqnarray}
\psi_\perp(r,\theta)=\sum_{s=1}^2~\gamma_s\left[A_s\left \{ \begin{array}{c}
             J_n(\alpha_sr) \\
             I_n(\alpha_sr)
           \end{array} \right\} + B_s \left \{ \begin{array}{c}
             Y_n(\alpha_sr) \\
             K_n(\alpha_sr)
           \end{array} \right\}\right] \Big[C_s\cos(n\theta)+D_s\sin(n\theta)\Big], \quad 
\end{eqnarray}
where
\begin{eqnarray}\label{solpsiPP2}
\displaystyle
\gamma_s=\left \{ \begin{array}{lr}
             1 & \text{if}~s=1 \\
             {1\over\kappa}\left(\kappa - {\rho\tau\over\mu}\right) & \text{if}~s=2
           \end{array} \right., 
\end{eqnarray}
\end{subequations}
and all other quantities are as defined for Eq.~(\ref{solphiP}). In fact, given $\phi_\perp(r,\theta)$ as expressed in Eq.~(\ref{phiperpGEN}), we can write 
\begin{equation}\label{solpsiPExp2}
\psi_\perp(r,\theta)=\sum_{s=1}^2\gamma_s\phi^{(s)}_r(r)\phi^{(s)}_\theta(\theta).
\end{equation}
Note that when $\displaystyle (\lambda+2\mu)\kappa = {\rho\tau}$, the correct expression for the $\phi^{(1)}_r(r)$ term in Eq.~(\ref{solpsiPExp2}) is given by Eq.~(\ref{LP1}). When $\displaystyle \mu\kappa = {\rho\tau}$, Eq.~(\ref{solpsiP}) reproduces the correct solution with the $s=2$ term vanishing identically. 

The only remaining task in solving system (\ref{systemeF}) is to determine the axial and temporal parts of the full potentials $\Phi$ and $\Psi$ [see Eq.~(\ref{solform})]. The conditions listed in (\ref{assump}) provide the governing equations for determining these functions. Conditions (\ref{assump3}), (\ref{assump1}), and (\ref{assump2}) yield
\begin{eqnarray}\label{Zpart}
\phi_z(z)=\psi_z(z)=\left \{ \begin{array}{lr}
             E\cos\left(\sqrt{|\kappa|}z\right)+F\sin\left(\sqrt{|\kappa|}z\right) & \text{if}~\kappa<0 \\
             E\exp\left(-\sqrt{\kappa}z\right)+F\exp\left(\sqrt{\kappa}z\right) & \text{if}~\kappa>0
           \end{array} \right.,
\end{eqnarray}
and similarly, conditions (\ref{assump6}), (\ref{assump4}), and (\ref{assump5}) yield
\begin{eqnarray}\label{Tpart}
\phi_t(t)=\psi_t(t)=\left \{ \begin{array}{lr}
             G\cos\left(\sqrt{|\tau|}t\right)+H\sin\left(\sqrt{|\tau|}t\right) & \text{if}~\tau<0 \\
             G\exp\left(-\sqrt{\tau}t\right)+H\exp\left(\sqrt{\tau}t\right) & \text{if}~\tau>0
           \end{array} \right.,
\end{eqnarray}
where $\{E,F,G,H\}$ are arbitrary constants.

Particular solutions to subsystem (\ref{systemeSFIN3}) can be obtained by assuming a product solution of the form
\begin{equation}\label{chiprodsol}
\chi(r,\theta,z,t)=\chi^{}_r(r)\chi^{}_\theta(\theta)\chi^{}_z(z)\chi^{}_t(t)
\end{equation}
and using separation of variables, as shown in Appendix \ref{AppSOV}. Although not necessary, more compact particular solutions to system (\ref{systemeSFIN}) can be obtained by identifying the separation constants $\{\eta_t,\eta_z, \eta_r\}$ (see Appendix \ref{AppSOV}) with the parameters $\kappa$ and $\tau$ as follows:
\begin{equation}\label{sepconsts}
\eta_z=\kappa, \quad  \quad \eta_t={\rho\tau\over\mu}, \quad  \quad \eta_r=\eta_t-\eta_z={\rho\tau\over\mu}-\kappa.
\end{equation}
Upon doing so, one can immediately conclude (from inspection of the solutions given in Appendix \ref{AppSOV}) that, apart from arbitrary constants, $\chi^{}_z(z)$ and $\chi^{}_t(t)$ are identical to $\phi_z(z)$ and $\phi_t(t)$, respectively, and that $\chi^{}_\perp(r,\theta)\equiv\chi^{}_r(r)\chi^{}_\theta(\theta)$ is (apart from arbitrary constants) identical to the $s=2$ term of Eq.~(\ref{solphiP}):
\begin{eqnarray}\label{solchiperp}
\chi^{}_\perp(r,\theta)=\left[
           A_3\left \{ \begin{array}{c}
             J_n(\alpha_2r) \\
             I_n(\alpha_2r)
           \end{array} \right\} + B_3 \left \{ \begin{array}{c}
             Y_n(\alpha_2r) \\
             K_n(\alpha_2r)
           \end{array} \right\}\right]\Big[C_3\cos(n\theta)+D_3\sin(n\theta)\Big], ~ \alpha_2\neq0             
\end{eqnarray}
\begin{eqnarray}\label{ZpartC}
\chi^{}_z(z)=\left \{ \begin{array}{lr}
             \widetilde{E}\cos\left(\sqrt{|\kappa|}z\right)+\widetilde{F}\sin\left(\sqrt{|\kappa|}z\right) & \text{if}~\kappa<0 \\
             \widetilde{E}\exp\left(-\sqrt{\kappa}z\right)+\widetilde{F}\exp\left(\sqrt{\kappa}z\right) & \text{if}~\kappa>0
           \end{array} \right.,
\end{eqnarray}
\begin{eqnarray}\label{TpartC}
\chi^{}_t(t)=\left \{ \begin{array}{lr}
             \widetilde{G}\cos\left(\sqrt{|\tau|}t\right)+\widetilde{H}\sin\left(\sqrt{|\tau|}t\right) & \text{if}~\tau<0 \\
             \widetilde{G}\exp\left(-\sqrt{\tau}t\right)+\widetilde{H}\exp\left(\sqrt{\tau}t\right) & \text{if}~\tau>0
           \end{array} \right.,
\end{eqnarray}
where $\displaystyle\left\{A_3,B_3,C_3,D_3,\widetilde{E},\widetilde{F},\widetilde{G},\widetilde{H}\right\}$ are arbitrary constants. Note that if $\alpha_2=0$, then the radial part of Eq.~(\ref{solchiperp}) should be replaced by
\begin{eqnarray}\label{solchirad2}
\chi^{}_r(r)=\left \{ \begin{array}{lr}
           A_3+B_3\ln r & ~\text{if}~n=0 \\ 
           A_3r^n+B_3r^{-n} & ~\text{if}~n\neq0 \\
           \end{array} \right., \quad \text{when}~\alpha_2=0.
\end{eqnarray}

Since all three Buchwald potentials have now been completely determined (under conditions (\ref{assump}) and (\ref{sepconsts})), we can write down the corresponding particular solutions to Eq.~(\ref{NLE}) using the component form of Eq.~(\ref{Buchy}):
\begin{equation}\label{Buchy22}
\mathbf{u}=u_r\mathbf{\hat{r}}+u_\theta{\boldsymbol{\hat{\theta}}}+u_z\mathbf{\hat{z}}=\left({\partial\Phi\over\partial r}+{1\over r}{\partial\chi\over\partial\theta}\right)\mathbf{\hat{r}} + \left({1\over r}{\partial\Phi\over\partial\theta}-{\partial\chi\over\partial r}\right){\boldsymbol{\hat{\theta}}}+{\partial\Psi\over\partial z}\mathbf{\hat{z}}. 
\end{equation}
Assuming $\displaystyle (\lambda+2\mu)\kappa \neq {\rho\tau}$ ($\alpha_1\neq0$) and $\displaystyle \mu\kappa \neq {\rho\tau}$ ($\alpha_2\neq0$), the cylindrical components of the displacement are then given by:  
\begin{eqnarray}\label{solnradcomp}
u_r&=&\left(\sum_{s=1}^2\left[A_s\left \{ \begin{array}{c}
             J_n'(\alpha_sr) \\
             I_n'(\alpha_sr)
           \end{array} \right\} + B_s \left \{ \begin{array}{c}
             Y_n'(\alpha_sr) \\
             K_n'(\alpha_sr)
           \end{array} \right\}\right] \Big[C_s\cos(n\theta)+D_s\sin(n\theta)\Big]\right)\phi_z(z)\phi_t(t) \nonumber \\ \nonumber \\
           &+&{n\over r}\left[A_3\left \{ \begin{array}{c}
             J_n(\alpha_2r) \\
             I_n(\alpha_2r)
           \end{array} \right\} + B_3 \left \{ \begin{array}{c}
             Y_n(\alpha_2r) \\
             K_n(\alpha_2r)
           \end{array} \right\}\right] \Big[-C_3\sin(n\theta)+D_3\cos(n\theta)\Big]\chi^{}_z(z)\chi^{}_t(t), \nonumber \\   
\end{eqnarray}
\begin{eqnarray}\label{solnangcomp}
u_\theta&=&{n\over r}\left(\sum_{s=1}^2\left[A_s\left \{ \begin{array}{c}
             J_n(\alpha_sr) \\
             I_n(\alpha_sr)
           \end{array} \right\} + B_s \left \{ \begin{array}{c}
             Y_n(\alpha_sr) \\
             K_n(\alpha_sr)
           \end{array} \right\}\right] \Big[-C_s\sin(n\theta)+D_s\cos(n\theta)\Big]\right)\phi_z(z)\phi_t(t) \nonumber \\ \nonumber \\
           &-&\left[A_3\left \{ \begin{array}{c}
             J_n'(\alpha_2r) \\
             I_n'(\alpha_2r)
           \end{array} \right\} + B_3 \left \{ \begin{array}{c}
             Y_n'(\alpha_2r) \\
             K_n'(\alpha_2r)
           \end{array} \right\}\right] \Big[C_3\cos(n\theta)+D_3\sin(n\theta)\Big]\chi^{}_z(z)\chi^{}_t(t), \nonumber \\   
\end{eqnarray}
and
\begin{eqnarray}\label{solnaxcomp}
u_z=\left(\sum_{s=1}^2~\gamma_s\left[A_s\left \{ \begin{array}{c}
             J_n(\alpha_sr) \\
             I_n(\alpha_sr)
           \end{array} \right\} + B_s \left \{ \begin{array}{c}
             Y_n(\alpha_sr) \\
             K_n(\alpha_sr)
           \end{array} \right\}\right] \Big[C_s\cos(n\theta)+D_s\sin(n\theta)\Big]\right){\text{d}\psi_z(z)\over\text{d}z}~\psi_t(t). \nonumber \\
\end{eqnarray}

\noindent The following should be noted in conjunction with Eqs.~(\ref{solnradcomp})-(\ref{solnaxcomp}): 

\noindent \textbf{(i)} The constant $\displaystyle n$ is a non-negative integer; 

\noindent \textbf{(ii)} The constants $\displaystyle \left\{A_1,A_2,A_3,B_1,B_2,B_3,C_1,C_2,C_3,D_1,D_2,D_3\right\}$ are arbitrary constants;

\noindent \textbf{(iii)} The constants $\alpha_1$ and $\alpha_2$ in the arguments of the Bessel functions are given by Eq.~(\ref{alphas});

\noindent \textbf{(iv)} The proper combination of Bessel functions is as given in Table \ref{TabLinCombos};

\noindent \textbf{(v)} In Eqs.~(\ref{solnradcomp})-(\ref{solnangcomp}), primes denote differentiation with respect to the radial coordinate $r$;

\noindent \textbf{(vi)} In Eqs.~(\ref{solnradcomp}) and (\ref{solnangcomp}), the functions $\phi_z(z)$, $\phi_t(t)$, $\chi^{}_z(z)$, and $\chi^{}_t(t)$ are given by Eqs.~(\ref{Zpart}), (\ref{Tpart}), (\ref{ZpartC}), and (\ref{TpartC}), respectively;

\noindent \textbf{(vii)} In Eq.~(\ref{solnaxcomp}), the functions $\psi_z(z)$ and $\psi_t(t)$ are given by Eqs.~(\ref{Zpart}) and (\ref{Tpart}), respectively, and the constant $\gamma_s$ is given by Eq.~(\ref{solpsiPP2});

\noindent \textbf{(viii)} Equations (\ref{solnradcomp})-(\ref{solnaxcomp}) are valid when $\alpha_1\neq0$ and $\alpha_2\neq0$; otherwise the radial parts must be modified as discussed in connection with Eqs.~(\ref{LP1}), (\ref{LP2}), and (\ref{solchirad2}). 

\subsection{Special Case: $\displaystyle \kappa=0$}

In deriving Eqs.~(\ref{solnradcomp})-(\ref{solnaxcomp}), we stipulated that the free parameter $\kappa$ defined in Eqs.~(\ref{assump1})-(\ref{assump2}), which subsequently appears in the definitions of the constants $\alpha_1$ and $\alpha_2$ [c.f., Eq.~(\ref{alphas})] and the constant $\gamma_2$ [c.f., Eq.~(\ref{solpsiPP2})], must be non-zero. Corresponding solutions can also be obtained when $\kappa=0$, as we will now show. In this special case, system (\ref{systemefin}) decouples and reduces to: 
\begin{subequations}\label{systemefinkapeq0} 
\begin{equation}\label{systemefin1kapeq0}
\left[(\lambda+2\mu)\nabla_\perp^2 - \rho\tau\right]\phi_\perp=0,
\end{equation}
\begin{equation}\label{systemefin2kapeq0}
\left[(\lambda+\mu)\nabla_\perp^2\right]\phi_\perp+\left[\mu\nabla_\perp^2-\rho\tau\right]\psi_\perp=0. 
\end{equation}
\end{subequations}
Equation (\ref{systemefin1kapeq0}) can be cast in the form of a 2D Helmholtz equation in polar coordinates and immediately solved using separation of variables. The following particular solutions to  Eq.~(\ref{systemefin1kapeq0}) (having $2\pi$-periodic angular parts) are then obtained: 
\begin{subequations}\label{solphiPkapeq0}
\begin{eqnarray}\label{solphiPkapeq0A}
\phi_\perp(r,\theta)=\left[A_1\left \{ \begin{array}{c}
             J_n(\alpha_1r) \\
             I_n(\alpha_1r)
           \end{array} \right\} + B_1 \left \{ \begin{array}{c}
             Y_n(\alpha_1r) \\
             K_n(\alpha_1r)
           \end{array} \right\}\right] \Big[C_1\cos(n\theta)+D_1\sin(n\theta)\Big],  
\end{eqnarray}
where
\begin{equation}\label{alphaskapeq0}
\alpha_1=\sqrt{~\left|{\rho\tau\over(\lambda+2\mu)}\right|~}.
\end{equation}
\end{subequations}
Note that the unmodified Bessel functions $\displaystyle\left\{J_n(\alpha_1r),Y_n(\alpha_1r)\right\}$ in the radial part of Eq.~(\ref{solphiPkapeq0A}) are to be used when $\displaystyle{\rho\tau/(\lambda+2\mu)}<0$ and the modified Bessel functions $\displaystyle\left\{I_n(\alpha_1r),K_n(\alpha_1r)\right\}$ when $\displaystyle{\rho\tau/(\lambda+2\mu)}>0$. 

To solve for $\psi_\perp(r,\theta)$, we substitute $(\lambda+\mu)\nabla_\perp^2\phi_\perp=-\left[\mu\nabla_\perp^2 - \rho\tau\right]\phi_\perp$, which is obtained from rearranging Eq.~(\ref{systemefin1kapeq0}), into Eq.~(\ref{systemefin2kapeq0}). This yields the PDE 
\begin{subequations}\label{Fperpeq} 
\begin{equation}\label{FperpeqA}
\left[\mu\nabla_\perp^2 - \rho\tau\right]\mathsf{W}_\perp=0,
\end{equation}
where
\begin{equation}\label{Fperpdef}
\mathsf{W}_\perp\equiv\left(\psi_\perp-\phi_\perp\right).
\end{equation}
\end{subequations}
As before, Eq.~(\ref{FperpeqA}) can be cast in the form of a 2D Helmholtz equation in polar coordinates and solved using separation of variables. This yields the following particular solutions (again having $2\pi$-periodic angular parts) to Eq.~(\ref{FperpeqA}): 
\begin{subequations}\label{solFperpeq}
\begin{eqnarray}\label{solFperpeqA}
\mathsf{W}_\perp(r,\theta)=\left[A_2\left \{ \begin{array}{c}
             J_n(\alpha_2r) \\
             I_n(\alpha_2r)
           \end{array} \right\} + B_2 \left \{ \begin{array}{c}
             Y_n(\alpha_2r) \\
             K_n(\alpha_2r)
           \end{array} \right\}\right] \Big[C_2\cos(n\theta)+D_2\sin(n\theta)\Big],  
\end{eqnarray}
where
\begin{equation}\label{alphaskapeq0P2}
\alpha_2=\sqrt{~{\rho\left|\tau\right|\over\mu}~}.
\end{equation}
\end{subequations}
The understanding here is that the unmodified Bessel functions $\displaystyle\left\{J_n(\alpha_2r),Y_n(\alpha_2r)\right\}$ in the radial part of Eq.~(\ref{solFperpeqA}) are to be used when $\displaystyle\tau<0$ and the modified Bessel functions $\displaystyle\left\{I_n(\alpha_2r),K_n(\alpha_2r)\right\}$ when $\displaystyle\tau>0$. By virtue of Eq.~(\ref{Fperpdef}), the solution for $\psi_\perp(r,\theta)$ is then given by
\begin{eqnarray}\label{solpsiperpkapeq0}
\psi_\perp(r,\theta)=\sum_{s=1}^2\left[A_s\left \{ \begin{array}{c}
             J_n(\alpha_sr) \\
             I_n(\alpha_sr)
           \end{array} \right\} + B_s \left \{ \begin{array}{c}
             Y_n(\alpha_sr) \\
             K_n(\alpha_sr)
           \end{array} \right\}\right] \Big[C_s\cos(n\theta)+D_s\sin(n\theta)\Big],  
\end{eqnarray}
where the constants $\alpha_1$ and $\alpha_2$ are given by Eqs.~(\ref{alphaskapeq0}) and (\ref{alphaskapeq0P2}), respectively. The correct linear combination of Bessel functions in the radial part of each term in Eq.~(\ref{solpsiperpkapeq0}) is as discussed above and summarized in Table \ref{TabLinComboskapeq0}. 

\vspace*{0.25cm}

\setlength{\extrarowheight}{10pt}
\begin{table}[h]
\centering
\begin{tabular}{| c | c | c |}
    \hline
    ~~~Linear Combination~~~ & ~~~~~ $s=1$ term ~~~~~ & ~~~~~$s=2$ term ~~~~~\\ [10pt] \hline
    $\displaystyle\{J_n(\alpha_sr), Y_n(\alpha_sr)\}$ & $\displaystyle {\rho\tau\over(\lambda+2\mu)}<0$ & $\displaystyle \tau<0$ \\ [10pt] \hline
    $\displaystyle\{I_n(\alpha_sr), K_n(\alpha_sr)\}$ & $\displaystyle {\rho\tau\over(\lambda+2\mu)}>0$ & $\displaystyle \tau>0$ \\ [10pt] \hline
\end{tabular}
\caption{Conditions on the radial part of each term in Eq.~(\ref{solpsiperpkapeq0}). By extension, the $s=1$ column also applies to the radial part of Eq.~(\ref{solphiPkapeq0A}) and the $s=2$ column to the radial part of Eq.~(\ref{solchiperpkapeq0}).}
\label{TabLinComboskapeq0}
\end{table}
\setlength{\extrarowheight}{1pt}

The axial parts of the full potentials $\Phi$ and $\Psi$ [c.f., Eq.~(\ref{solform})], as determined from conditions (\ref{assump3}), (\ref{assump1}), and (\ref{assump2}) are 
\begin{equation}\label{Zpartkapeq0}
\phi_z(z)=\psi_z(z)=E+Fx, 
\end{equation}
where $\{E,F\}$ are arbitrary constants. The temporal parts of potentials $\Phi$ and $\Psi$ are unchanged [see Eq.~(\ref{Tpart})]. 

At this point, the only outstanding ingredient is the Buchwald potential $\chi$, which is governed by Eq.~(\ref{systemeSFIN3}), and by assumption, has the separable form given in Eq.~(\ref{chiprodsol}). Under conditions (\ref{sepconsts}), one can immediately conclude from inspection of the solutions given in Appendix \ref{AppSOV} that, apart from arbitrary constants, the components $\chi^{}_z(z)$ and $\chi^{}_t(t)$ [c.f., Eq.~(\ref{chiprodsol})] are identical to $\phi_z(z)$ and $\phi_t(t)$ as given by Eqs.~(\ref{Zpartkapeq0}) and (\ref{Tpart}), respectively, and that $\chi^{}_\perp(r,\theta)\equiv\chi^{}_r(r)\chi^{}_\theta(\theta)$ is (apart from arbitrary constants) identical to the function $\mathsf{W}_\perp(r,\theta)$ given by Eq.~(\ref{solFperpeqA}). Thus, analogous to Eqs.~(\ref{solchiperp})-(\ref{ZpartC}), the spatial components of $\chi$ [c.f., Eq.~(\ref{chiprodsol})] can be expressed as follows:  
\begin{eqnarray}\label{solchiperpkapeq0}
\chi^{}_\perp(r,\theta)=\left[
           A_3\left \{ \begin{array}{c}
             J_n(\alpha_2r) \\
             I_n(\alpha_2r)
           \end{array} \right\} + B_3 \left \{ \begin{array}{c}
             Y_n(\alpha_2r) \\
             K_n(\alpha_2r)
           \end{array} \right\}\right]\Big[C_3\cos(n\theta)+D_3\sin(n\theta)\Big],            
\end{eqnarray}
where $\alpha_2$ is given by Eq.~(\ref{alphaskapeq0P2}), and 
\begin{equation}\label{ZpartCkapeq0}
\chi^{}_z(z)=\widetilde{E}+\widetilde{F}x. 
\end{equation}
The temporal part of $\chi$ (i.e., $\chi^{}_t(t)$) is given by Eq.~(\ref{TpartC}). As in Eqs.~(\ref{solchiperp})-(\ref{ZpartC}), the constants $\displaystyle\left\{A_3,B_3,C_3,D_3,\widetilde{E},\widetilde{F}\right\}$ appearing in the above equations are arbitrary constants. 

With all three Buchwald potentials now completely determined (under the stipulated conditions), we can write down the corresponding particular solutions to Eq.~(\ref{NLE}) using (\ref{Buchy22}):
\begin{eqnarray}\label{solnradcompkapeq0}
u_r&=&\left[A_1\left \{ \begin{array}{c}
             J_n'(\alpha_1r) \\
             I_n'(\alpha_1r)
           \end{array} \right\} + B_1 \left \{ \begin{array}{c}
             Y_n'(\alpha_1r) \\
             K_n'(\alpha_1r)
           \end{array} \right\}\right] \Big[C_1\cos(n\theta)+D_1\sin(n\theta)\Big]\phi_z(z)\phi_t(t) \nonumber \\ \nonumber \\
           &+&{n\over r}\left[A_3\left \{ \begin{array}{c}
             J_n(\alpha_2r) \\
             I_n(\alpha_2r)
           \end{array} \right\} + B_3 \left \{ \begin{array}{c}
             Y_n(\alpha_2r) \\
             K_n(\alpha_2r)
           \end{array} \right\}\right] \Big[-C_3\sin(n\theta)+D_3\cos(n\theta)\Big]\chi^{}_z(z)\chi^{}_t(t), \nonumber \\ 
\end{eqnarray}
\begin{eqnarray}\label{solnangcompkapeq0}
u_\theta&=&{n\over r}\left[A_1\left \{ \begin{array}{c}
             J_n(\alpha_1r) \\
             I_n(\alpha_1r)
           \end{array} \right\} + B_1 \left \{ \begin{array}{c}
             Y_n(\alpha_1r) \\
             K_n(\alpha_1r)
           \end{array} \right\}\right] \Big[-C_1\sin(n\theta)+D_1\cos(n\theta)\Big]\phi_z(z)\phi_t(t) \nonumber \\ \nonumber \\
           &-&\left[A_3\left \{ \begin{array}{c}
             J_n'(\alpha_2r) \\
             I_n'(\alpha_2r)
           \end{array} \right\} + B_3 \left \{ \begin{array}{c}
             Y_n'(\alpha_2r) \\
             K_n'(\alpha_2r)
           \end{array} \right\}\right] \Big[C_3\cos(n\theta)+D_3\sin(n\theta)\Big]\chi^{}_z(z)\chi^{}_t(t), \nonumber \\   
\end{eqnarray}
and
\begin{eqnarray}\label{solnaxcompkapeq0}
u_z=\left(\sum_{s=1}^2\left[A_s\left \{ \begin{array}{c}
             J_n(\alpha_sr) \\
             I_n(\alpha_sr)
           \end{array} \right\} + B_s \left \{ \begin{array}{c}
             Y_n(\alpha_sr) \\
             K_n(\alpha_sr)
           \end{array} \right\}\right] \Big[C_s\cos(n\theta)+D_s\sin(n\theta)\Big]\right){\text{d}\psi_z(z)\over\text{d}z}~\psi_t(t). \nonumber \\
\end{eqnarray}

\noindent In conjunction with Eqs.~(\ref{solnradcompkapeq0})-(\ref{solnaxcompkapeq0}), the following should be noted: 

\noindent \textbf{(N)} As in Eqs.~(\ref{solnradcomp})-(\ref{solnaxcomp}), $\left\{A_1,A_2,A_3,B_1,B_2,B_3,C_1,C_2,C_3,D_1,D_2,D_3\right\}$ are arbitrary constants, and $\displaystyle n$ is a non-negative integer; 

\noindent \textbf{(i)} The constants $\alpha_1$ and $\alpha_2$ in the arguments of the Bessel functions are given by Eqs.~(\ref{alphaskapeq0}) and (\ref{alphaskapeq0P2}), respectively;

\noindent \textbf{(ii)} The proper combination of Bessel functions is as given in Table \ref{TabLinComboskapeq0};

\noindent \textbf{(iii)} In Eqs.~(\ref{solnradcompkapeq0})-(\ref{solnangcompkapeq0}), primes denote differentiation with respect to the radial coordinate $r$;

\noindent \textbf{(iv)} In Eqs.~(\ref{solnradcompkapeq0}) and (\ref{solnangcompkapeq0}), the functions $\phi_z(z)$, $\phi_t(t)$, $\chi^{}_z(z)$, and $\chi^{}_t(t)$ are given by Eqs.~(\ref{Zpartkapeq0}), (\ref{Tpart}), (\ref{ZpartCkapeq0}), and (\ref{TpartC}), respectively;

\noindent \textbf{(v)} In Eq.~(\ref{solnaxcompkapeq0}), the functions $\psi_z(z)$ and $\psi_t(t)$ are given by Eqs.~(\ref{Zpartkapeq0}) and (\ref{Tpart}), respectively;

\noindent \textbf{(vi)} Equations (\ref{solnradcompkapeq0})-(\ref{solnaxcompkapeq0}) are only valid when the free parameter $\kappa$ defined in Eqs.~(\ref{assump1})-(\ref{assump2}) is zero.  

\section{Applications}\label{AppPrologue}

Our focus thus far has been on solving the PDE system (\ref{systemeSFIN}) under assumptions (\ref{solform},\ref{chiprodsol}) and conditions (\ref{assump},\ref{sepconsts}) and then using (\ref{Buchy22}) to arrive at Eqs.~(\ref{solnradcomp})-(\ref{solnaxcomp}) and Eqs.~(\ref{solnradcompkapeq0})-(\ref{solnaxcompkapeq0}). Independent of how the latter two sets of equations were actually derived, we have at our disposal a set of particular solutions to the NL equation that are interesting and useful in their own right. These parametric solutions can be directly applied to a fundamental set of linear elastic boundary-value problems (BVPs) in cylindrical coordinates, in particular, problems wherein the displacement, stress, and strain fields are of a prescribed form: constant or sinusoidal in the circumferential coordinate; constant, linear, sinusoidal, exponential, or hyperbolic in the longitudinal coordinate; and sinusoidal or exponential in the time coordinate. In the remaining sections of the paper, we shall illustrate through examples how Eqs.~(\ref{solnradcomp})-(\ref{solnaxcomp}) can be used to construct solutions to such BVPs. The BVPs we consider in this paper serve as a good starting point insofar as they are mathematically non-trivial but not overly complicated. The examples will also serve to highlight a key strength of using Eqs.~(\ref{solnradcomp})-(\ref{solnaxcomp}) to construct solutions to BVPs (from the aforementioned fundamental set of BVPs), which is that they are especially useful for constructing solutions to BVPs involving one or more physical parameters. Solutions to more interesting and mathematically complex BVPs can also be constructed with the aid of Eqs.~(\ref{solnradcomp})-(\ref{solnaxcomp}) and Eqs.~(\ref{solnradcompkapeq0})-(\ref{solnaxcompkapeq0}), but it is not within the scope of the present paper to do so. We hope to report on more advanced applications elsewhere.  

\section{Application 1: Forced Vibration of a Solid Cylinder}\label{EGFULL}

In this section, we demonstrate how the parametric solutions obtained in Sec.~\ref{PartSolns} can be used to efficiently construct exact solutions to an axisymmetric forced-vibration problem. Consider the problem of a solid elastic cylinder of finite (but arbitrary) length $L$ and radius $R$ subjected to a time-harmonic surface pressure that is constant circumferentially but varies along its axis according to a cosine law. (The cylinder is assumed to have a circular cross-section and to be homogeneous and isotropic.) The flat ends of the cylinder are clamped in the axial direction with zero shear. If the flat ends of the cylinder are situated at $z=0$ and $z=L$, then the boundary conditions for this axisymmetric problem are:
\begin{subequations}\label{BCsEG}
\begin{equation}\label{strssrr}
\sigma_{rr}(R,z,t)=\mathcal{A}\cos\left({2k\pi\over L}z\right)\sin(\omega t),
\end{equation}
\begin{equation}\label{strssrz}
\sigma_{rz}(R,z,t)=0,
\end{equation}
\begin{equation}\label{SSBCs1}
\sigma_{rz}(r,0,t)=\sigma_{rz}(r,L,t)=0,
\end{equation}
\begin{equation}\label{SSBCs2}
u_z(r,0,t)=u_z(r,L,t)=0,
\end{equation}
\end{subequations}
where $\sigma_{rr}(r,z,t)$ is a normal component of stress, $\sigma_{rz}(r,z,t)$ is a shear component of stress, $\mathcal{A}$ and $k$ are prescribed constants, and $\omega$ is the prescribed angular frequency of excitation. Note that $\mathcal{A}\in\mathbb{R}\backslash\{0\}$ is a constant stress having units of pressure and $k\in\mathbb{Z}^+$ is dimensionless. Conditions (\ref{strssrr}) and (\ref{strssrz}) are pure stress boundary conditions on the curved surface of the cylinder; they must be satisfied for all $z\in(0,L)$ and arbitrary $t$. Conditions (\ref{SSBCs1}) and (\ref{SSBCs2}) are admissible mixed boundary conditions specifying orthogonal components of the stress and displacement at the flat ends of the cylinder \cite{LittleChilds67}; they must be satisfied for all $r\in[0,R]$ and arbitrary $t$. The condition that the displacement field is finite everywhere in the cylinder is implicit. 

Note that we have not given any information about the displacement field and its time derivatives at time $t=t_0$, and thus the problem defined above is not an initial-value problem. We seek to determine the \emph{steady state} vibration response of the cylinder when it is subjected to the non-uniform distribution of stress (\ref{strssrr}) on its curved surface. Since the excitation is axisymmetric (i.e., independent of the $\theta$ coordinate) and time-harmonic, the response of the cylinder must be so as well. We may thus immediately conclude that the displacement, stress, and strain fields are all independent of the $\theta$ coordinate, and furthermore, that $u_\theta=0$. 

The axial and temporal parts of the displacement field can be readily obtained with the aid of the stress-displacement relations from linear elasticity. For an axisymmetric problem, these are given by \cite{VTCS2010}:
\begin{subequations}\label{strssdispCYLaxi}
\begin{equation}\label{stssstrnCYL1axi}
\sigma_{rr}=(\lambda+2\mu){\partial u_r\over\partial r}+{\lambda\over r}u_r+\lambda{\partial u_z\over\partial z},
\end{equation}
\begin{equation}\label{stssstrnCYL2axi}
\sigma_{\theta\theta}=\lambda{\partial u_r\over\partial r}+{(\lambda+2\mu)\over r}u_r+\lambda{\partial u_z\over\partial z},
\end{equation}
\begin{equation}\label{stssstrnCYL3axi}
\sigma_{zz}=\lambda{\partial u_r\over\partial r}+{\lambda\over r}u_r+(\lambda+2\mu){\partial u_z\over\partial z},
\end{equation}
\begin{equation}\label{stssstrnCYL5axi}
\sigma_{rz}=\mu\left({\partial u_r\over\partial z}+{\partial u_z\over\partial r}\right)=\sigma_{zr}.
\end{equation}
\end{subequations}
Let $\kappa=-\left({2k\pi\over L}\right)^2$ and $F=0$ in Eq.~(\ref{Zpart}), $\tau=-\omega^2$ and $G=0$ in Eq.~(\ref{Tpart}), and $n=0$ in Eqs.~(\ref{solnradcomp}) and (\ref{solnaxcomp}). By then comparing Eqs.~(\ref{solnradcomp}), (\ref{solnaxcomp}), (\ref{stssstrnCYL1axi}), and (\ref{strssrr}), we may immediately deduce that the axial and temporal parts of the displacement components are given by 
\begin{equation}\label{phipsiZEG}
\phi_z(z)=\psi_z(z)=E\cos\left({2k\pi\over L}z\right),
\end{equation}
\begin{equation}\label{phipsiTEG}
\phi_t(t)=\psi_t(t)=H\sin(\omega t).
\end{equation}
The radial parts of the displacement components depend on the relative values of the material and excitation parameters; five cases can be distinguished as listed in Table \ref{EG1Cases}. The problem must be solved separately for each of these five cases. \emph{In the following, we shall provide complete solutions for only Cases 1 and 4.} Cases 1 and 4 are regular and degenerate cases, respectively, and are representative of the two types of cases. The remaining cases can be solved similarly, and we shall provide only the final results. 

\setlength{\extrarowheight}{10pt}
\begin{table}[h]
\centering
\begin{tabular}{| c | c | c |} \hline 
    ~Case~ & ~Parametric Relationship ($k$, $\omega$)~ & ~Parametric Relationship ($\kappa$, $\tau$)~ \\ [5pt] \hline 
     1 & $\displaystyle {\rho\omega^2\over(\lambda+2\mu)}<{\rho\omega^2\over\mu}<\left({2k\pi\over L}\right)^2$ & $\displaystyle \kappa<{\rho\tau\over(\lambda+2\mu)}$ and $\displaystyle \kappa<{\rho\tau\over\mu}$ \\ [10pt] \hline 
     2 & $\displaystyle {\rho\omega^2\over(\lambda+2\mu)}<{\rho\omega^2\over\mu}=\left({2k\pi\over L}\right)^2$ & $\displaystyle \kappa<{\rho\tau\over(\lambda+2\mu)}$ and $\displaystyle \kappa={\rho\tau\over\mu}$ \\ [10pt] \hline 
     3 & $\displaystyle {\rho\omega^2\over(\lambda+2\mu)}<\left({2k\pi\over L}\right)^2<{\rho\omega^2\over\mu}$ & $\displaystyle \kappa<{\rho\tau\over(\lambda+2\mu)}$ and $\displaystyle \kappa>{\rho\tau\over\mu}$ \\ [10pt] \hline 
     4 & $\displaystyle \left({2k\pi\over L}\right)^2={\rho\omega^2\over(\lambda+2\mu)}<{\rho\omega^2\over\mu}$ & $\displaystyle \kappa={\rho\tau\over(\lambda+2\mu)}$ and $\displaystyle \kappa>{\rho\tau\over\mu}$ \\ [10pt] \hline 
     5 & $\displaystyle \left({2k\pi\over L}\right)^2<{\rho\omega^2\over(\lambda+2\mu)}<{\rho\omega^2\over\mu}$ & $\displaystyle \kappa>{\rho\tau\over(\lambda+2\mu)}$ and $\displaystyle \kappa>{\rho\tau\over\mu}$ \\ [10pt] \hline 
\end{tabular}
\caption{Parametric relationships defining five distinct sub-problems. In the second column, the relationship is expressed in terms of the physical excitation parameters $k$ and $\omega$, whereas in the third column, the relationship is expressed in terms of the mathematical parameters $\kappa$ and $\tau$. Note the implicit assumption that $\lambda>0$.}
\label{EG1Cases}
\end{table} 
\setlength{\extrarowheight}{1pt}

\subsection{\textbf{Case 1:} $\displaystyle{\rho\omega^2\over(\lambda+2\mu)}<{\rho\omega^2\over\mu}<\left({2k\pi\over L}\right)^2$}

In this case, $\kappa<{\rho\tau/(\lambda+2\mu)}$ and $\kappa<{\rho\tau/\mu}$ (c.f., Table \ref{EG1Cases}), and thus, according to Table \ref{TabLinCombos}, linear combinations of $\displaystyle\{I_0(\alpha_sr), K_0(\alpha_sr)\}$ (and their derivatives) should be employed in the radial parts of Eqs.~(\ref{solnradcomp}) and (\ref{solnaxcomp}), where the constants $\alpha_1$ and $\alpha_2$, as determined from Eq.~(\ref{alphas}), are:
\begin{equation}\label{alphasEG}
\alpha_1=\sqrt{\left({2k\pi\over L}\right)^2-{\rho\omega^2\over(\lambda+2\mu)}}, \quad \alpha_2=\sqrt{\left({2k\pi\over L}\right)^2-{\rho\omega^2\over\mu}}.
\end{equation}
Since the modified Bessel functions of the second kind $K_p(\alpha_sr)\to\infty$ as $r\to0$ ($p\ge0$), such radial terms should be discarded, dictating that the constants $B_1$ and $B_2$ in the radial parts of Eqs.~(\ref{solnradcomp}) and (\ref{solnaxcomp}) are both zero. The constant $\gamma_s$ in Eq.~(\ref{solnaxcomp}), as determined from Eq.~(\ref{solpsiPP2}), is given by 
\begin{eqnarray}\label{solpsiPP2EG}
\displaystyle
\gamma_s=\left \{ \begin{array}{lr}
             1 & ~~\text{if}~s=1 \\
             1-\left[{\rho\omega^2\over\left({2k\pi\over L}\right)^2\mu}\right] & ~~\text{if}~s=2
           \end{array} \right..
\end{eqnarray}
Inputting all of the above ingredients into Eqs.~(\ref{solnradcomp}) and (\ref{solnaxcomp}), taking the necessary derivatives, and defining a new set of arbitrary constants $\bar{a}_s\equiv A_sC_sEH$, the displacement components take the form
\begin{equation}\label{RcompEG}
u_r(r,z,t)=\sum_{s=1}^2\bar{a}_s\alpha_sI_1(\alpha_sr)\cos\left({2k\pi\over L}z\right)\sin(\omega t),
\end{equation} 
\begin{equation}\label{ZcompEG}
u_z(r,z,t)=-\left({2k\pi\over L}\right)\sum_{s=1}^2\bar{a}_s\gamma_sI_0(\alpha_sr)\sin\left({2k\pi\over L}z\right)\sin(\omega t),
\end{equation}
where the constants $\alpha_s$ and $\gamma_s$ are given by Eqs.~(\ref{alphasEG}) and (\ref{solpsiPP2EG}), respectively. Note that $u_z(r,z,t)$ as given in Eq.~(\ref{ZcompEG}) automatically satisfies boundary condition (\ref{SSBCs2}). 

To complete the solution, we need only to determine the values of the constants $\bar{a}_1$ and $\bar{a}_2$ in Eqs.~(\ref{RcompEG}) and (\ref{ZcompEG}) that satisfy boundary conditions (\ref{strssrr}) and (\ref{strssrz}). Substituting Eqs.~(\ref{RcompEG}) and (\ref{ZcompEG}) into Eqs.~(\ref{stssstrnCYL1axi}) and (\ref{stssstrnCYL5axi}) and performing the necessary algebra yields the stress components
\begin{subequations}\label{strsscompEGRR}
\begin{equation}\label{strsscompEGgenRR1}
\sigma_{rr}(r,z,t)=\Bigg[\sum_{s=1}^2\bar{a}_s\left[\beta_sI_0(\alpha_sr)-{2\mu\alpha_s\over r}I_1(\alpha_sr)\right]\Bigg]\cos\left({2k\pi\over L}z\right)\sin(\omega t),
\end{equation}
where
\begin{equation}\label{strsscompEGgenRR2}
\beta_s=\lambda\left[\alpha^2_s-\gamma_s\left({2k\pi\over L}\right)^2\right]+2\mu\alpha^2_s,
\end{equation}
\end{subequations}
and 
\begin{equation}\label{strsscompEGgenRZ}
\sigma_{rz}(r,z,t)=-\mu\left({2k\pi\over L}\right)\Bigg[\sum_{s=1}^2\bar{a}_s\alpha_s\left(1+\gamma_s\right)I_1(\alpha_sr)\Bigg]\sin\left({2k\pi\over L}z\right)\sin(\omega t).
\end{equation}
Note that $\sigma_{rz}(r,z,t)$ as given by Eq.~(\ref{strsscompEGgenRZ}) automatically satisfies boundary condition (\ref{SSBCs1}). 
Substituting Eq.~(\ref{strsscompEGRR}) into the LHS of boundary condition (\ref{strssrr}) and then canceling the $\cos\left({2k\pi\over L}z\right)\sin(\omega t)$ terms on both sides of the resulting equation yields the condition 
\begin{subequations}\label{BCsEGsmplfd}
\begin{equation}\label{BCsEGsmplfdRR}
\sum_{s=1}^2\bar{a}_s\left[\beta_sI_0(\alpha_sR)-{2\mu\alpha_s\over R}I_1(\alpha_sR)\right]=\mathcal{A}.
\end{equation}
Using (\ref{strsscompEGgenRZ}), boundary condition (\ref{strssrz}) can be written as
\begin{equation}\label{BCsEGsmplfdRZ1}
-\mu\left({2k\pi\over L}\right)\Bigg[\sum_{s=1}^2\bar{a}_s\alpha_s\left(1+\gamma_s\right)I_1(\alpha_sR)\Bigg]\sin\left({2k\pi\over L}z\right)\sin(\omega t)=0,
\end{equation}
which must be satisfied for all $z\in(0,L)$ and arbitrary $t$ thus implying the condition 
\begin{equation}\label{BCsEGsmplfdRZ2}
\sum_{s=1}^2\bar{a}_s\alpha_s\left(1+\gamma_s\right)I_1(\alpha_sR)=0.
\end{equation}
\end{subequations}
Conditions (\ref{BCsEGsmplfdRR}) and (\ref{BCsEGsmplfdRZ2}) can be written as the $2\times2$ linear system
\begin{eqnarray}\label{BCmatrixEQ}
\left[\begin{array}{cc}
            \beta_1I_0(\alpha_1R)-{2\mu\alpha_1\over R}I_1(\alpha_1R) ~&~ \beta_2I_0(\alpha_2R)-{2\mu\alpha_2\over R}I_1(\alpha_2R) \\
             \alpha_1\left(1+\gamma_1\right)I_1(\alpha_1R) & \alpha_2\left(1+\gamma_2\right)I_1(\alpha_2R)
           \end{array} \right] \left[\begin{array}{c}
             \bar{a}_1 \\
             \bar{a}_2
           \end{array} \right] =  \left[\begin{array}{c}
              \mathcal{A} \\
             0
           \end{array} \right],
\end{eqnarray}
whose solution may be expressed (after some tedious but straightforward algebra) in the following simplified form:
\begin{subequations}\label{solncoeffsEG}
\begin{equation}\label{solncoeffsEG1}
\bar{a}_1={\alpha_2\left(1+\gamma_2\right)I_1(\alpha_2R)\mathcal{A}\over2\mu\alpha_1\alpha_2\Upsilon(\alpha_1,\alpha_2,\gamma_1,\gamma_2,R)+\lambda\left(\alpha^2_1-\left({2k\pi\over L}\right)^2\right)\alpha_2(1+\gamma_2)I_0(\alpha_1R)I_1(\alpha_2R)},
\end{equation}
\begin{equation}\label{solncoeffsEG2}
\bar{a}_2=-{\alpha_1\left(1+\gamma_1\right)I_1(\alpha_1R)\mathcal{A}\over2\mu\alpha_1\alpha_2\Upsilon(\alpha_1,\alpha_2,\gamma_1,\gamma_2,R)+\lambda\left(\alpha^2_1-\left({2k\pi\over L}\right)^2\right)\alpha_2(1+\gamma_2)I_0(\alpha_1R)I_1(\alpha_2R)},
\end{equation}
where
\begin{equation}\label{solncoeffsEG3}
\Upsilon(\alpha_1,\alpha_2,\gamma_1,\gamma_2,R)=(1+\gamma_2)I'_1(\alpha_1R)I_1(\alpha_2R)-(1+\gamma_1)I_1(\alpha_1R)I'_1(\alpha_2R),
\end{equation}
and
\begin{equation}\label{solncoeffsEG4}
I'_1(\alpha_sR)=\alpha_sI_0(\alpha_sR)-{I_1(\alpha_sR)\over R},~\quad s=1,2.
\end{equation}
\end{subequations}
The components of the displacement field are thus given by Eqs.~(\ref{RcompEG}) and (\ref{ZcompEG}) with constants $\bar{a}_s$ given by Eq.~(\ref{solncoeffsEG}).  

\subsection{\textbf{Case 2:} $\displaystyle {\rho\omega^2\over\mu}=\left({2k\pi\over L}\right)^2$}\label{degencase2}

An analysis parallel to that given for Case 4 below shows that Eqs.~(\ref{alphasEG})-(\ref{strsscompEGgenRZ}) with $\alpha_2=0$ and 
\begin{eqnarray}\label{solpsiPP2EGC2}
\gamma_s=\left \{ \begin{array}{lr}
             1 & ~~\text{if}~s=1 \\
             0 & ~~\text{if}~s=2
           \end{array} \right.
\end{eqnarray} 
reproduce the correct general expressions for the displacement and stress components in this case. The analogs of Eqs.~(\ref{BCsEGsmplfdRR}) and (\ref{BCsEGsmplfdRZ2}) that arise from application of boundary conditions (\ref{strssrr}) and (\ref{strssrz}) are in this case given by
\begin{subequations}\label{BCsEGsmplfdC2}
\begin{equation}\label{BCsEGsmplfdRRC2}
\bar{a}_1\left[\beta_1I_0(\alpha_1R)-{2\mu\alpha_1\over R}I_1(\alpha_1R)\right]=\mathcal{A},
\end{equation}
\begin{equation}\label{BCsEGsmplfdRZC2}
\bar{a}_1I_1(\alpha_1R)=0.
\end{equation}
\end{subequations}
Since $I_1(x)=0$ implies that $x=0$, Eq.~(\ref{BCsEGsmplfdRZC2}) can be satisfied only if: (i) $\alpha_1=0$ or (ii) $\bar{a}_1=0$. Option (i) may be immediately ruled out since $\alpha_1\neq0$. Thus,  satisfying Eq.~(\ref{BCsEGsmplfdRZC2}) requires $\bar{a}_1=0$. However, $\bar{a}_1=0$ cannot satisfy Eq.~(\ref{BCsEGsmplfdRRC2}) because of the restriction $\mathcal{A}\neq0$. Thus, no consistent solution to system (\ref{BCsEGsmplfdC2}) exists when $\mathcal{A}\neq0$. Our general parametric formulas for the displacement field therefore yield no solution in this degenerate case. If the restriction $\mathcal{A}\neq0$ is removed, then the trivial solution $u_r(r,z,t)=u_z(r,z,t)=0$ is a consistent solution in the force-free (i.e., $\mathcal{A}=0$) case. 

\subsection{\textbf{Case 3:} $\displaystyle {\rho\omega^2\over(\lambda+2\mu)}<\left({2k\pi\over L}\right)^2<{\rho\omega^2\over\mu}$}

In this case, the displacement components take the form
\begin{equation}\label{RcompEG3}
u_r(r,z,t)=\bigg[\bar{a}_1\alpha_1I_1(\alpha_1r)-\bar{a}_2\alpha_2J_1(\alpha_2r)\bigg]\cos\left({2k\pi\over L}z\right)\sin(\omega t),
\end{equation} 
\begin{equation}\label{ZcompEG3}
u_z(r,z,t)=-\left({2k\pi\over L}\right)\bigg[\bar{a}_1\gamma_1I_0(\alpha_1r)+\bar{a}_2\gamma_2J_0(\alpha_2r)\bigg]\sin\left({2k\pi\over L}z\right)\sin(\omega t),
\end{equation}
where 
\begin{equation}\label{alphasEG3}
\alpha_1=\sqrt{\left({2k\pi\over L}\right)^2-{\rho\omega^2\over(\lambda+2\mu)}}, \quad \alpha_2=\sqrt{-\left({2k\pi\over L}\right)^2+{\rho\omega^2\over\mu}},
\end{equation}
and $\gamma_s$ ($s=1,2$) is again given by Eq.~(\ref{solpsiPP2EG}). The constants $\bar{a}_1$ and $\bar{a}_2$ appearing in Eqs.~(\ref{RcompEG3}) and (\ref{ZcompEG3}) are again determined from application of boundary conditions (\ref{strssrr}) and (\ref{strssrz}); their values in this case are as follows:
\begin{subequations}\label{solcoeffsEG3}
\begin{equation}\label{solcoeffsEG31}
\bar{a}_1={\alpha_2\left(1+\gamma_2\right)J_1(\alpha_2R)\mathcal{A}\over2\mu\alpha_1\alpha_2\Upsilon(\alpha_1,\alpha_2,\gamma_1,\gamma_2,R)+\lambda\left(\alpha^2_1-\left({2k\pi\over L}\right)^2\right)\alpha_2(1+\gamma_2)I_0(\alpha_1R)J_1(\alpha_2R)},
\end{equation}
\begin{equation}\label{solcoeffsEG32}
\bar{a}_2={\alpha_1\left(1+\gamma_1\right)I_1(\alpha_1R)\mathcal{A}\over2\mu\alpha_1\alpha_2\Upsilon(\alpha_1,\alpha_2,\gamma_1,\gamma_2,R)+\lambda\left(\alpha^2_1-\left({2k\pi\over L}\right)^2\right)\alpha_2(1+\gamma_2)I_0(\alpha_1R)J_1(\alpha_2R)},
\end{equation}
where
\begin{equation}\label{solcoeffsEG33}
\Upsilon(\alpha_1,\alpha_2,\gamma_1,\gamma_2,R)=(1+\gamma_2)I'_1(\alpha_1R)J_1(\alpha_2R)-(1+\gamma_1)I_1(\alpha_1R)J'_1(\alpha_2R),
\end{equation}
and
\begin{equation}\label{solcoeffsEG34}
I'_1(\alpha_1R)=\alpha_1I_0(\alpha_1R)-{I_1(\alpha_1R)\over R},~\quad J'_1(\alpha_2R)=\alpha_2J_0(\alpha_2R)-{J_1(\alpha_2R)\over R}.
\end{equation}
\end{subequations}

\subsection{\textbf{Case 4:} $\displaystyle \left({2k\pi\over L}\right)^2={\rho\omega^2\over(\lambda+2\mu)}$}\label{degencase4}

In this case, the constant $\alpha_1=0$, and as described in comment \textbf{(viii)} at the end of Section \ref{solnkapneq0}, the radial parts of Eqs.~(\ref{solnradcomp}) and (\ref{solnaxcomp}) must be carefully modified. First recall that $n=0$ in Eqs.~(\ref{solnradcomp}) and (\ref{solnaxcomp}) since the problem is axisymmetric, and thus according to Eq.~(\ref{LP1}), the $s=1$ term in each of the transverse parts of  Eqs.~(\ref{solnradcomp}) and (\ref{solnaxcomp}) should employ linear combinations of $\{1,\ln r\}$ (and their derivatives) rather than linear combinations of Bessel functions. The $s=2$ terms in the transverse parts of Eqs.~(\ref{solnradcomp}) and (\ref{solnaxcomp}) do not require any modification, and according to Table \ref{TabLinCombos}, linear combinations of $\displaystyle\{J_0(\alpha_2r), Y_0(\alpha_2r)\}$ (and their derivatives) should be employed in this case, where the constant $\alpha_2$, as determined from Eq.~(\ref{alphas}), is given by 
\begin{equation}\label{alpha2EG2}
\alpha_2=\sqrt{-\left({2k\pi\over L}\right)^2+{\rho\omega^2\over\mu}}. 
\end{equation}
By defining a new set of arbitrary constants $\bar{a}_s\equiv A_sC_sEH$ and $\bar{b}_s\equiv B_sC_sEH$ ($s=1,2$), the displacement components immediately take the following general forms: 
\begin{eqnarray}\label{RcompEG2}
u_r(r,z,t)&=&{\text{d}\over\text{d}r}\Bigg(\Big[\bar{a}_1+\bar{b}_1\ln r\Big]+\Big[\bar{a}_2J_0(\alpha_2r)+\bar{b}_2Y_0(\alpha_2r)\Big]\Bigg)\cos\left({2k\pi\over L}z\right)\sin(\omega t) \nonumber \\
&=&\left({\bar{b}_1\over r}-\alpha_2\Big[\bar{a}_2J_1(\alpha_2r)+\bar{b}_2Y_1(\alpha_2r)\Big]\right)\cos\left({2k\pi\over L}z\right)\sin(\omega t),
\end{eqnarray}
\begin{equation}\label{ZcompEG2GEN}
u_z(r,z,t)=-\left({2k\pi\over L}\right)\Bigg(\gamma_1\Big[\bar{a}_1+\bar{b}_1\ln r\Big]+\gamma_2\Big[\bar{a}_2J_0(\alpha_2r)+\bar{b}_2Y_0(\alpha_2r)\Big]\Bigg)\sin\left({2k\pi\over L}z\right)\sin(\omega t),
\end{equation}
where $\gamma_s$ ($s=1,2$) is again given by Eq.~(\ref{solpsiPP2EG}), which in this special case reduces to
\begin{eqnarray}\label{solpsiPP2EG2R}
\gamma_s=\left \{ \begin{array}{lr}
             1 & ~~\text{if}~s=1 \\
             1-{(\lambda+2\mu)\over\mu} & ~~\text{if}~s=2
           \end{array} \right..
\end{eqnarray}
Since $Y_p(\alpha_2r)\to-\infty$ as $r\to0$ ($p\ge0$) and $\ln r\to-\infty$ as $r\to0$, these terms (as well as the ${1/r}$ term) must be discarded, dictating that $\bar{b}_1=0$ and $\bar{b}_2=0$ in Eqs.~(\ref{RcompEG2}) and (\ref{ZcompEG2GEN}). The displacement components thus reduce to:
\begin{equation}\label{RcompEG2R}
u_r(r,z,t)=-\bar{a}_2\alpha_2J_1(\alpha_2r)\cos\left({2k\pi\over L}z\right)\sin(\omega t),
\end{equation}
\begin{equation}\label{ZcompEG2R}
u_z(r,z,t)=-\left({2k\pi\over L}\right)\Bigg(\bar{a}_1\gamma_1+\bar{a}_2\gamma_2J_0(\alpha_2r)\Bigg)\sin\left({2k\pi\over L}z\right)\sin(\omega t).
\end{equation} 
As before, $u_z(r,z,t)$ satisfies boundary condition (\ref{SSBCs2}). 

All that remains is to determine the values of the constants $\bar{a}_1$ and $\bar{a}_2$ in Eqs.~(\ref{RcompEG2R}) and (\ref{ZcompEG2R}) that satisfy boundary conditions (\ref{strssrr}) and (\ref{strssrz}). Substituting Eqs.~(\ref{RcompEG2R}) and (\ref{ZcompEG2R}) into Eqs.~(\ref{stssstrnCYL1axi}) and (\ref{stssstrnCYL5axi}) and performing the necessary algebra yields the stress components
\begin{equation}\label{strsscompEG2genRR}
\sigma_{rr}(r,z,t)=-\Bigg[\bar{a}_1\lambda\left({2k\pi\over L}\right)^2+\bar{a}_2\bigg[2\mu\alpha_2^2J_0(\alpha_2r)-{2\mu\alpha_2\over r}J_1(\alpha_2r)\bigg]\Bigg]\cos\left({2k\pi\over L}z\right)\sin(\omega t),
\end{equation}
and 
\begin{equation}\label{strsscompEG2RgenRZ}
\sigma_{rz}(r,z,t)=\mu\left({2k\pi\over L}\right)\bar{a}_2\alpha_2\left(1+\gamma_2\right)J_1(\alpha_2r)\sin\left({2k\pi\over L}z\right)\sin(\omega t).
\end{equation}
Note that $\sigma_{rz}(r,z,t)$ as given by Eq.~(\ref{strsscompEG2RgenRZ}) automatically satisfies boundary condition (\ref{SSBCs1}). 
Substituting Eq.~(\ref{strsscompEG2genRR}) into the LHS of boundary condition (\ref{strssrr}) and then canceling the $\cos\left({2k\pi\over L}z\right)\sin(\omega t)$ terms on both sides of the resulting equation yields the condition 
\begin{subequations}\label{BCsEG2smplfd}
\begin{equation}\label{BCsEG2smplfdRR}
-\Bigg[\bar{a}_1\lambda\left({2k\pi\over L}\right)^2+\bar{a}_2\bigg[2\mu\alpha_2^2J_0(\alpha_2R)-{2\mu\alpha_2\over R}J_1(\alpha_2R)\bigg]\Bigg]=\mathcal{A}.
\end{equation}
Using (\ref{strsscompEG2RgenRZ}), boundary condition (\ref{strssrz}) can be written as
\begin{equation}\label{BCsEG2RsmplfdRZ1}
\bar{a}_2\Bigg[\mu\left({2k\pi\over L}\right)\alpha_2\left(1+\gamma_2\right)J_1(\alpha_2R)\Bigg]\sin\left({2k\pi\over L}z\right)\sin(\omega t)=0,
\end{equation}
which must be satisfied for all $z\in[0,L]$ and arbitrary $t$ thus implying the condition 
\begin{equation}\label{BCsEG2smplfdRZ2}
\bar{a}_2\left(1+\gamma_2\right)J_1(\alpha_2R)=0.
\end{equation}
\end{subequations}
Conditions (\ref{BCsEG2smplfdRR}) and (\ref{BCsEG2smplfdRZ2}) can be written as the $2\times2$ linear system
\begin{eqnarray}\label{BCmatrixEQ2}
\left[\begin{array}{cc}
            -\lambda \left({2k\pi\over L}\right)^2 ~&~ -2\mu\alpha_2^2J_0(\alpha_2R)+{2\mu\alpha_2\over R}J_1(\alpha_2R) \\
             0 & \left(1+\gamma_2\right)J_1(\alpha_2R)
           \end{array} \right] \left[\begin{array}{c}
             \bar{a}_1 \\
             \bar{a}_2
           \end{array} \right] =  \left[\begin{array}{c}
              \mathcal{A} \\
             0
           \end{array} \right].
\end{eqnarray}
A unique solution to system (\ref{BCmatrixEQ2}) exists only when $\gamma_2\neq-1$ \emph{and} $J_1(\alpha_2R)\neq0$, otherwise the system is underdetermined and there exist an infinite number of solutions. It can be readily inferred from Eq.~(\ref{solpsiPP2EG2R}) that $\gamma_2=-1$ corresponds to the mathematically permissible but physically unusual situation in which $\lambda$=0. Although solutions for this unusual case can be readily obtained, we regard these solutions to be extraneous in the context of classical linear elasticity. In this subsection, we shall therefore impose the restriction $\gamma_2\neq-1$.    

\subsubsection{\textbf{Case 4(i):} $\gamma_2\neq-1$ and $J_1(\alpha_2R)\neq0$}

In this sub-case, the unique solution to system (\ref{BCmatrixEQ2}) is easily found; the result is: 
\begin{equation}\label{case41soln}
\bar{a}_1=-{\mathcal{A}\over\lambda\left({2k\pi\over L}\right)^2}, ~~\bar{a}_2=0.
\end{equation}
The components of the displacement field are therefore
\begin{equation}\label{finREG2}
u_r(r,z,t)=0,
\end{equation}
\begin{equation}\label{finZEG2}
u_z(r,z,t)={\mathcal{A}\over\lambda\left({2k\pi\over L}\right)}\sin\left({2k\pi\over L}z\right)\sin(\omega t).
\end{equation}

\subsubsection{\textbf{Case 4(ii):} $\gamma_2\neq-1$ and $J_1(\alpha_2R)=0$}\label{degensoln2} 

In this sub-case, the infinitude of solutions to system (\ref{BCmatrixEQ2}) is given by 
\begin{subequations}\label{case42solnG}
\begin{equation}\label{case42solnG1}
\bar{a}_1=-{1\over\lambda\left({2k\pi\over L}\right)^2}\bigg[\mathcal{A}+2\mu\bar{a}_2\alpha_2^2J_0(\alpha_2R)\bigg],
\end{equation}
\begin{equation}\label{case42solnG2}
\bar{a}_2=C\in\mathbb{R}\backslash\{0\}.
\end{equation}
\end{subequations}
(Note that the free parameter $C$ has dimensions of $(\text{length})^2$.) The components of the displacement field are therefore
\begin{equation}\label{finREG242}
u_r(r,z,t)=-C\alpha_2J_1(\alpha_2r)\cos\left({2k\pi\over L}z\right)\sin(\omega t),
\end{equation}
\begin{equation}\label{finZEG242}
u_z(r,z,t)={1\over\left({2k\pi\over L}\right)}\Bigg({\mathcal{A}\over\lambda}+C\alpha^2_2\left[{2\mu J_0(\alpha_2R)\over\lambda}+J_0(\alpha_2r)\right]\Bigg)\sin\left({2k\pi\over L}z\right)\sin(\omega t).
\end{equation} 
The constant $\alpha_2$ in this sub-case is such that $\alpha_2=q_m/R$, where $q_m$ is the $m$th positive zero of $J_1(x)$, and $m\in\mathbb{Z}^+$. In other words, the excitation parameters $k$ and $\omega$ simultaneously satisfy the relations 
\begin{equation}\label{excitrelations}
\left({2k\pi\over L}\right)^2={\rho\omega^2\over(\lambda+2\mu)}, \quad \quad \sqrt{-\left({2k\pi\over L}\right)^2+{\rho\omega^2\over\mu}~}={q_m\over R}.
\end{equation} 
Thus, whenever the excitation and material parameters are prescribed in such a way that relations (\ref{excitrelations}) are simultaneously satisfied, there exist a continuous family of solutions [Eqs.~(\ref{finREG242}) and (\ref{finZEG242})] parametrized by the free parameter $C$. 

\subsection{\textbf{Case 5:} $\displaystyle \left({2k\pi\over L}\right)^2<{\rho\omega^2\over(\lambda+2\mu)}<{\rho\omega^2\over\mu}$}

In this case, the displacement components take the form
\begin{equation}\label{RcompEG5}
u_r(r,z,t)=-\Bigg[\sum_{s=1}^2\bar{a}_s\alpha_sJ_1(\alpha_sr)\Bigg]\cos\left({2k\pi\over L}z\right)\sin(\omega t),
\end{equation} 
\begin{equation}\label{ZcompEG5}
u_z(r,z,t)=-\left({2k\pi\over L}\right)\Bigg[\sum_{s=1}^2\bar{a}_s\gamma_sJ_0(\alpha_sr)\Bigg]\sin\left({2k\pi\over L}z\right)\sin(\omega t),
\end{equation}
where 
\begin{equation}\label{alphasEG5}
\alpha_1=\sqrt{-\left({2k\pi\over L}\right)^2+{\rho\omega^2\over(\lambda+2\mu)}}, \quad \alpha_2=\sqrt{-\left({2k\pi\over L}\right)^2+{\rho\omega^2\over\mu}},
\end{equation}
and $\gamma_s$ ($s=1,2$) is again given by Eq.~(\ref{solpsiPP2EG}). The constants $\bar{a}_1$ and $\bar{a}_2$ appearing in Eqs.~(\ref{RcompEG5}) and (\ref{ZcompEG5}), as determined from application of boundary conditions (\ref{strssrr}) and (\ref{strssrz}), are as follows:
\begin{subequations}\label{coeffsEG5}
\begin{equation}\label{coeffsEG51}
\bar{a}_1=-{\alpha_2\left(1+\gamma_2\right)J_1(\alpha_2R)\mathcal{A}\over2\mu\alpha_1\alpha_2\Upsilon(\alpha_1,\alpha_2,\gamma_1,\gamma_2,R)+\lambda\left(\alpha^2_1+\left({2k\pi\over L}\right)^2\right)\alpha_2(1+\gamma_2)J_0(\alpha_1R)J_1(\alpha_2R)},
\end{equation}
\begin{equation}\label{coeffsEG52}
\bar{a}_2={\alpha_1\left(1+\gamma_1\right)J_1(\alpha_1R)\mathcal{A}\over2\mu\alpha_1\alpha_2\Upsilon(\alpha_1,\alpha_2,\gamma_1,\gamma_2,R)+\lambda\left(\alpha^2_1+\left({2k\pi\over L}\right)^2\right)\alpha_2(1+\gamma_2)J_0(\alpha_1R)J_1(\alpha_2R)},
\end{equation}
where
\begin{equation}\label{coeffsEG53}
\Upsilon(\alpha_1,\alpha_2,\gamma_1,\gamma_2,R)=(1+\gamma_2)J'_1(\alpha_1R)J_1(\alpha_2R)-(1+\gamma_1)J_1(\alpha_1R)J'_1(\alpha_2R),
\end{equation}
and
\begin{equation}\label{coeffsEG54}
J'_1(\alpha_sR)=\alpha_sJ_0(\alpha_sR)-{J_1(\alpha_sR)\over R},~\quad s=1,2.
\end{equation}
\end{subequations}

In closing this section, we mention that, once the displacement field is known, one can proceed to easily compute other quantities of interest, such as the strain field. For example, using the axisymmetric strain-displacement equations \cite{VTCS2010}
\begin{equation}\label{strnaxisym}
\varepsilon_{rr}(r,z,t)={\partial u_r(r,z,t)\over\partial r},~~~\varepsilon_{\theta\theta}(r,z,t)={u_r(r,z,t)\over r},~~~\varepsilon_{zz}(r,z,t)={\partial u_z(r,z,t)\over\partial z},
\end{equation}
one can immediately obtain the normal components of the strain field, and the corresponding volumetric strain, denoted here by $\mathcal{E}(r,z,t)$. For Case 1, the result is: 
\begin{equation}\label{VolStrn}
\mathcal{E}(r,z,t)\equiv\varepsilon_{rr}+\varepsilon_{\theta\theta}+\varepsilon_{zz}=\Bigg[\sum_{s=1}^2\bar{a}_s\left(\alpha^2_s-\gamma_s\left({2k\pi\over L}\right)^2\right)I_0(\alpha_sr)\Bigg]\cos\left({2k\pi\over L}z\right)\sin(\omega t), 
\end{equation}
where the constants $\bar{a}_s$ are again given by Eq.~(\ref{solncoeffsEG}). The shear strain(s) could also just as easily be calculated.

As a check, we have verified that, in all cases, the displacement field satisfies the NL equation [Eq.~(\ref{NLE})] and the boundary conditions (\ref{BCsEG}). 

\section{Application 2: Forced Relaxation of a Solid Cylinder}\label{rlxnEG}

To give the reader some appreciation of the range of applicability of the obtained parametric solutions, we briefly consider another class of problems, namely those in which a stressed solid is relaxed over time. The key feature that we wish to emphasize in this brief section is that the obtained parametric solutions can be applied to problems where the longitudinal disturbances are not sinusoidal. The solutions obtained in Sec.~\ref{PartSolns} can also be applied to problems where the longitudinal disturbances are constant, linear, exponential or hyperbolic. Since the emphasis here is on the nature of the longitudinal disturbances, we shall again restrict attention to axisymmetric problems. The other main feature illustrated by consideration of these problems is that the temporal dependence need not be harmonic. The obtained parametric solutions can also be applied to problems where the temporal dependence is exponential in nature.    

As an example of a hyperbolic axial disturbance, consider again a solid elastic cylinder of length $L$ and radius $R$ with the following pure stress boundary conditions on the curved surface
\begin{subequations}\label{BCsBEG1}
\begin{equation}\label{strssrrrzBEG1}
\sigma_{rr}(R,z,t)=\mathcal{A}\cosh\left({k\over L}z\right){\displaystyle b^{-{c\over T}t}}, \quad \sigma_{rz}(R,z,t)=0,
\end{equation}
where $\mathcal{A}\neq0$, and \emph{one} of the following unmixed or mixed boundary conditions prescribed at the flat ends of the cylinder (at $z=0,L$)
\begin{equation}\label{SBCsBEG1s1}
u_r(r,0,t)=u_r(r,L,t)=0, ~u_z(r,0,t)=0, ~u_z(r,L,t)=u_1{\displaystyle b^{-{c\over T}t}}, 
\end{equation}
\begin{equation}\label{SBCsBEG1s2}
\sigma_{rz}(r,0,t)=0, ~\sigma_{rz}(r,L,t)=p_1{\displaystyle b^{-{c\over T}t}}, ~\sigma_{zz}(r,0,t)=p_2{\displaystyle b^{-{c\over T}t}}, ~\sigma_{zz}(r,L,t)=p_3{\displaystyle b^{-{c\over T}t}}, 
\end{equation}
\begin{equation}\label{SBCsBEG1s3}
u_r(r,0,t)=u_r(r,L,t)=0, ~\sigma_{zz}(r,0,t)=p_2{\displaystyle b^{-{c\over T}t}}, ~\sigma_{zz}(r,L,t)=p_3{\displaystyle b^{-{c\over T}t}},
\end{equation}
\begin{equation}\label{SBCsBEG1s4}
u_z(r,0,t)=\sigma_{rz}(r,0,t)=0, ~u_z(r,L,t)=u_1{\displaystyle b^{-{c\over T}t}}, ~\sigma_{rz}(r,L,t)=p_1{\displaystyle b^{-{c\over T}t}},
\end{equation}
\end{subequations}
where $\sigma_{rr}(r,z,t)$ and $\sigma_{zz}(r,z,t)$ are normal components of stress, $\sigma_{rz}(r,z,t)$ is a shear component of stress, $\{\mathcal{A},p_1,p_2,p_3\}$ are prescribed constant stresses, $u_1$ is a prescribed constant displacement, $\{k>0,~b>1,~c>0\}$ are prescribed dimensionless constants, and $T$ is a prescribed relaxation time, that is, the time at which the radial stress at the fixed end ($z=0$) of the cylinder is $b^{-c}\mathcal{A}$. It is implicitly assumed that $t\ge0$, that is, the relaxation of the solid commences at $t=0$ \footnote{Note that this is not an initial-value problem since the displacement and/or stress fields (and their time derivatives) are not specified at $t=0$.}. A primary objective in such problems is to understand how the stationary part of the displacement field varies with respect to the relaxation time $T$. 

The stress parameters will generally be such that $(\lambda+2\mu)T^2\neq\rho\big[(c\ln b)(L/k)\big]^2$, in which case, closed-form solutions to the above-specified boundary-value problems cannot be obtained. If however the parameters are prescribed so that $(\lambda+2\mu)T^2=\rho\big[(c\ln b)(L/k)\big]^2$, an analysis that closely parallels that given for Case 4 of Sec.~\ref{EGFULL} yields the following conditional solution \footnote{The analysis proceeds from identifying the mathematical parameters $\{\kappa,\tau\}$ with the physical parameters $\{k,L,b,c,T\}$ as follows: $\kappa=(k/L)^2>0$ and $\tau=(c\ln b/T)^2>0$. It then follows that $\phi_z(z)=\psi_z(z)=E\cosh(kz/L)$ and $\phi_t(t)=\psi_t(t)=G\exp[-(c\ln b)t/T]=Gb^{-ct/T}$. Note that the inequality $(k/L)^2<\rho(c\ln b/T)^2/\mu$ must hold in this special case.}:
\begin{equation}\label{EG2P4}
u_r(r,z,t)=0, \quad u_z(r,z,t)={\mathcal{A}L\over\lambda k}\sinh\left({k\over L}z\right){\displaystyle b^{-{c\over T}t}},
\end{equation}
\begin{equation}\label{EG2P4P2}
\sigma_{rr}=\mathcal{A}\cosh\left({k\over L}z\right){\displaystyle b^{-{c\over T}t}}, \quad \sigma_{rz}=0, \quad \sigma_{zz}={(\lambda+2\mu)\over\lambda}\mathcal{A}\cosh\left({k\over L}z\right){\displaystyle b^{-{c\over T}t}}.
\end{equation}
The condition(s) under which the above displacement and stress fields are a solution (in this special case) are given in Table \ref{condsolns}. Note that the non-zero component of the displacement field can in this case be written as
\begin{equation}\label{EG2P4sec}
u_z(r,z,t)=A(T)\sinh\left({k\over L}z\right){\displaystyle b^{-{c\over T}t}}=\mathcal{A}\sqrt{{(\lambda+2\mu)\over\rho}}{T\over\lambda(c\ln b)}\sinh\left({k\over L}z\right){\displaystyle b^{-{c\over T}t}},
\end{equation}
from which one can clearly discern a linear dependence of the amplitude of $u_z(r,z,t)$ on the relaxation time $T$ (i.e., $A(T)\sim T$).

\begin{table}[h]
\centering
\begin{tabular}{| c | c |}
    \hline
    ~~~Boundary Value Problem~~~ & ~~~Condition(s) for a Solution~~~ \\ \hline
    (\ref{strssrrrzBEG1}) + (\ref{SBCsBEG1s1}) & $u_1=\mathcal{A}L\sinh(k)/\lambda k$  \\ \hline
    (\ref{strssrrrzBEG1}) + (\ref{SBCsBEG1s2}) & ~~$p_1=0,~p_2=(\lambda+2\mu)\mathcal{A}/\lambda,~p_3=(\lambda+2\mu)\mathcal{A}\cosh(k)/\lambda$~~  \\ \hline
    (\ref{strssrrrzBEG1}) + (\ref{SBCsBEG1s3}) & $p_2=(\lambda+2\mu)\mathcal{A}/\lambda,~p_3=(\lambda+2\mu)\mathcal{A}\cosh(k)/\lambda$  \\ \hline
    (\ref{strssrrrzBEG1}) + (\ref{SBCsBEG1s4}) & $u_1=\mathcal{A}L\sinh(k)/\lambda k,~p_1=0$  \\ \hline
\end{tabular}
\caption{Conditions under which Eqs.~(\ref{EG2P4}) and (\ref{EG2P4P2}) are a solution to the boundary value problems defined by Eqs.~(\ref{strssrrrzBEG1})-(\ref{SBCsBEG1s4}). These solutions are valid only when the parameters are such that $(\lambda+2\mu)T^2=\rho\big[(c\ln b)(L/k)\big]^2$.}
\label{condsolns}
\end{table}

In closing this short section, we must emphasize that solution (\ref{EG2P4}) could not have been obtained without due consideration of the zero roots of Eq.~(\ref{chareq}). 

\section{Discussion}

As we have discussed in Sec.~\ref{AppPrologue}, the utility of having parametric solutions to the NL equation (like Eqs.~(\ref{solnradcomp})-(\ref{solnaxcomp})) is that they can used to efficiently construct analytical solutions to certain types of linear elastic BVPs. To highlight the beneficial features of our obtained parametric solutions and the advantages of using them for analytical work, it is useful to compare our method of solving the axisymmetric forced-vibration problem studied in Sec.~\ref{EGFULL} with other applicable methods available in the literature. Although laborious, it can be shown that the formalism of Ebenezer et al.~\cite{Ebenezer05} (henceforth referred to as ERP) when applied to the BVP defined by Eqs.~(\ref{BCsEG}) yields the same set of solutions that we obtained in Sec.~\ref{EGFULL}, with the exception of the degenerate solution obtained in Sec.~\ref{degensoln2}, which cannot be extracted from ERP's formalism. Note however that ERP's formalism does not make explicit use of modified Bessel functions and so the analytical formulas of Ref.~\cite{Ebenezer05} are much less convenient since one has to deal separately with the issue of imaginary prefactors and imaginary arguments occurring in the Bessel functions that appear in ERP's formulas.  

Our method of solving BVP (\ref{BCsEG}) may (at first sight) appear to be more complicated than applying ERP's formalism to the same problem. One must however keep in mind that we have explicitly considered all possible parameter regimes involving the excitation frequency. All of these separate cases are implicit in ERP's formalism but they need to be unraveled by hand in order to obtain explicit analytical solutions corresponding to all of the various cases. So, applying ERP's method may superficially appear to be simpler than our method, but as far as obtaining an explicit analytical solution to BVP (\ref{BCsEG}) is concerned, ERP's method is no less labor intensive. The advantage in using our approach to analytically solve forced-vibration problems like the one defined by Eqs.~(\ref{BCsEG})   is that the unraveling of the various cases and their corresponding solutions has already been achieved and it is just a matter of identifying the physical parameters of the problem with the mathematical parameters contained in the parametric solutions. 

The biggest advantages of using the approach demonstrated in this paper lie beyond the confines of the example considered in Sec.~\ref{EGFULL}. The parametric solutions obtained in Sec.~\ref{PartSolns} can also be applied to \emph{non-axisymmetric} forced-vibration problems involving solid or hollow elastic cylinders, as well as open cylindrical panels, with various boundary conditions, and to problems involving transient responses that may, for instance, be modeled by power laws in time. Forced-relaxation problems, such as the one considered in Sec.~\ref{rlxnEG}, are another class of problems that can be studied using the demonstrated approach. As a point of clarification, it should be emphasized that none of the aforementioned problems can be studied using the formalism of Ref.~\cite{Ebenezer05} or its extensions \cite{Ebenezer08,Ebenezer15}. 

\section{Conclusion}

Using a separable Buchwald representation in cylindrical coordinates, we have shown how under certain conditions the coupled equations of motion governing the Buchwald potentials can be decoupled and then solved using well-known techniques from the theory of PDEs. Under these conditions, we then constructed three parametrized families of particular solutions to the NL equation. Note that the obtained solutions possess $2\pi$-periodic angular parts. Solutions lacking this property can also be constructed and we hope to elaborate on this matter elsewhere. Our method of solution is much more general than the conventional cylindrical-wave substitution method that is ubiquitous in the literature and has the capacity to be further generalized. 

To our knowledge, the parametric solutions obtained here are the first of their kind in the sense that analogous parametric solutions have not yet been derived from any of the other well-known  representations. Analogous parametric solutions that derive from the Helmholtz-Lam\'{e} decomposition, for instance, are not available. Such parametric solutions can, in principle, be obtained, but for problems formulated in cylindrical coordinates, it is not clear whether the effort is justified given that the Buchwald representation is most analytically efficient in this system of coordinates.     

The parametric solutions obtained in this paper can be used to efficiently solve a fundamental set of linear elastic boundary value problems in cylindrical coordinates, in particular, problems wherein the displacement, stress, and strain fields are of a prescribed form, namely: constant or sinusoidal in the circumferential coordinate; and constant, linear, sinusoidal, exponential, or hyperbolic in the longitudinal coordinate. Time-harmonic solutions, which are relevant to a variety of problems, including forced-vibration problems, can be obtained by simply setting the parameter $\tau=-\omega^2$, where $\omega$ is the angular frequency. Equations (\ref{Tpart}) and (\ref{TpartC}) show that exponential time dependences also satisfy the NL equation. Analytical solutions having such temporal functions are useful for solving certain types of forced-relaxation problems and for modeling transient elastic motions that decay exponentially with time. 

As an illustrative application, we considered the forced-response problem for an axially-clamped solid elastic cylinder subjected to an axisymmetric harmonic standing-wave excitation on its curved surface, and using the family of parametric solutions that are sinusoidal in the axial coordinate, we constructed exact solutions for the displacement field in all allowed parameter regimes. Besides the intrinsic efficiency of the Buchwald decomposition for solving problems formulated in cylindrical coordinates, which is well-documented in the literature, the main advantage of using the parametric solutions obtained here is that, for problems involving one or more physical parameters, one can identify all physically allowed parameter regimes and obtain solutions valid in each of these regimes using a single comprehensive formalism. In future, we hope to apply the ideas and results of this paper to more complex forced-motion problems, especially asymmetric problems. 

This work is part of a larger effort whose purpose is to find effective analytical methods for generating parametrized families of particular solutions to the NL equation in cylindrical coordinates that can be usefully applied to modeling the elastic motions of finite cylindrical structures. As part of this larger effort, we hope to extend the present work in a future publication by addressing the problem of finding analytical solutions to Eq.~(\ref{NLBE}) without reverting to the sufficient condition given by Eq.~(\ref{systemeB}).

\section*{Acknowledgments}
The authors acknowledge financial support from the Natural Sciences and Engineering Research Council (NSERC) of Canada and the Ontario Research Foundation (ORF). 

\appendix

\section{Solution to Subsystem (\ref{systemeSFIN3}) using Separation of Variables}\label{AppSOV}

\noindent Substituting (\ref{chiprodsol}) into Eq.~(\ref{systemeSFIN3}) and then dividing the result by $\chi^{}_r(r)\chi^{}_\theta(\theta)\chi^{}_z(z)\chi^{}_t(t)$ yields
\begin{equation}\label{SOVEq}
\left[{\chi_r''(r)\over\chi^{}_r(r)} + {1\over r}{\chi_r'(r)\over\chi^{}_r(r)} + {1\over r^2} {\chi_\theta''(\theta)\over\chi^{}_\theta(\theta)}\right] + \left[{\chi_z''(z)\over\chi^{}_z(z)}\right] = \left[{1\over c_T^2}{\chi_t''(t)\over\chi^{}_t(t)}\right],
\end{equation}
where $\displaystyle c_T\equiv\sqrt{\mu/\rho}$. The first bracketed term depends only on $r$ and $\theta$, the second bracketed term only on $z$, and the right-hand side depends only on $t$. The above equation can be satisfied for all values of $r$, $\theta$, $z$, and $t$ only if each of the bracketed terms is a constant. This leads to the equations
\begin{equation}\label{Teq}
\chi_t''(t)-\eta_tc_T^2\chi^{}_t(t)=0,
\end{equation}
\begin{equation}\label{Zeq}
\chi_z''(z)-\eta_z\chi^{}_z(z)=0,
\end{equation}
and
\begin{equation}\label{RandTHeq}
\left[{\chi_r''(r)\over\chi^{}_r(r)} + {1\over r}{\chi_r'(r)\over\chi^{}_r(r)} + {1\over r^2} {\chi_\theta''(\theta)\over\chi^{}_\theta(\theta)}\right]=\eta_r,
\end{equation}
where the separation constants $\eta_t$, $\eta_z$, and $\eta_r$ obey (by virtue of Eq.~(\ref{SOVEq})) the relation
\begin{equation}\label{constsrel}
\eta_r+\eta_z=\eta_t.
\end{equation}
Equation (\ref{RandTHeq}) can be written in the form
\begin{equation}\label{RandTHeq2}
r^2{\chi_r''(r)\over\chi^{}_r(r)} + r{\chi_r'(r)\over\chi^{}_r(r)} -r^2\eta_r = -{\chi_\theta''(\theta)\over\chi^{}_\theta(\theta)}.
\end{equation}
The left-hand side of Eq.~(\ref{RandTHeq2}) depends only on $r$ and the right-hand side only on $\theta$. The two sides must therefore be equal to another constant, which we shall denote by $\eta_\theta$. This second separation of variables yields the equations
\begin{equation}\label{DEchiradpart}
r^2\chi_r''(r) +r\chi_r'(r) -\left(r^2\eta_r+\eta_\theta\right)\chi^{}_r(r)=0,
\end{equation}
\begin{equation}\label{DEchiangpart}
\chi_\theta''(\theta)+\eta_\theta\chi^{}_\theta(\theta)=0.
\end{equation}
Note that all real values of the separation constants $\eta_\theta$, $\eta_t$, $\eta_z$, and $\eta_r$ are admissible but that the latter three constants are not all independent. Upon independent specification of any two of $\{\eta_t,\eta_z, \eta_r\}$, the third is automatically determined by relation (\ref{constsrel}). Without loss of generality, we may assign $\eta_t$ and $\eta_z$ to be independent free parameters (that can take on any real values) whereupon $\eta_r=\eta_t-\eta_z$.
Particular solutions to Eqs.~(\ref{DEchiangpart}), (\ref{Zeq}), and (\ref{Teq}) are then  
\begin{eqnarray}\label{chiangpart}
\chi^{}_\theta(\theta)=\left \{ \begin{array}{lr}
             \widetilde{C}\exp\left(-\sqrt{|\eta_\theta|}\theta\right)+\widetilde{D}\exp\left(\sqrt{|\eta_\theta|}\theta\right) & \text{if}~\eta_\theta<0 \\
             \widetilde{C} + \widetilde{D}\theta & \text{if}~\eta_\theta=0 \\
             \widetilde{C}\cos\left(\sqrt{\eta_\theta}\theta\right)+\widetilde{D}\sin\left(\sqrt{\eta_\theta}\theta\right) & \text{if}~\eta_\theta>0
           \end{array} \right.,
\end{eqnarray}
\begin{eqnarray}\label{chiZsol}
\chi^{}_z(z)=\left \{ \begin{array}{lr}
             \widetilde{E}\cos\left(\sqrt{|\eta_z|}z\right)+\widetilde{F}\sin\left(\sqrt{|\eta_z|}z\right) & \text{if}~\eta_z<0 \\
             \widetilde{E} + \widetilde{F}z & \text{if}~\eta_z=0 \\
             \widetilde{E}\exp\left(-\sqrt{\eta_z}z\right)+\widetilde{F}\exp\left(\sqrt{\eta_z}z\right) & \text{if}~\eta_z>0
           \end{array} \right.,
\end{eqnarray}  
and
\begin{eqnarray}\label{chiTsol}
\chi^{}_t(t)=\left \{ \begin{array}{lr}
             \widetilde{G}\cos\left(\sqrt{|\eta_t|}c_Tt\right)+\widetilde{H}\sin\left(\sqrt{|\eta_t|}c_Tt\right) & \text{if}~\eta_t<0 \\
             \widetilde{G} + \widetilde{H}t & \text{if}~\eta_t=0 \\
             \widetilde{G}\exp\left(-\sqrt{\eta_t}c_Tt\right)+\widetilde{H}\exp\left(\sqrt{\eta_t}c_Tt\right) & \text{if}~\eta_t>0
           \end{array} \right.,
\end{eqnarray}  
respectively. If $\chi^{}_\theta(\theta)$ is to be a $2\pi$-periodic function, then 
\begin{equation}\label{etathetaP}
\eta_\theta=n^2, \quad n=0,1,2,3,\ldots, 
\end{equation}
in which case $\chi^{}_\theta(\theta)$ may be written as
\begin{equation}\label{chithetasoln}
\chi^{}_\theta(\theta)=\widetilde{C}\cos\left(n\theta\right)+\widetilde{D}\sin\left(n\theta\right), \quad n=0,1,2,3,\ldots. 
\end{equation}
Particular solutions to Eq.~(\ref{DEchiradpart}), given $\eta_r=\eta_t-\eta_z$, are then 
\begin{eqnarray}\label{chiradsol}
\chi^{}_r(r)=\left \{ \begin{array}{lr}
           \widetilde{A}J_n\left(\sqrt{|\eta_r|}r\right)+\widetilde{B}Y_n\left(\sqrt{|\eta_r|}r\right) & \text{if}~\eta_r<0 \\
           \widetilde{A}+\widetilde{B}\ln r & \text{if}~\eta_r=0~\text{and}~n=0 \\ 
           \widetilde{A}r^n+\widetilde{B}r^{-n} & \text{if}~\eta_r=0~\text{and}~n\neq0 \\
           \widetilde{A}I_n\left(\sqrt{\eta_r}r\right)+\widetilde{B}K_n\left(\sqrt{\eta_r}r\right) & \text{if}~\eta_r>0
           \end{array} \right..
\end{eqnarray}

\end{document}